\documentclass{article}
\usepackage[margin=1in]{geometry} % full-width
\usepackage{graphicx} % Required for inserting images
\usepackage{epsfig,amsmath,amssymb,amsfonts}
\usepackage{xcolor}
\usepackage{cite}

\def\Real{\mathbb{R}}
\advance\parskip 2.5pt
\def\Re{\mathop{\rm Re}\nolimits}
\def\Im{\mathop{\rm Im}\nolimits}

\catcode`\@ 11
\let\tru@int=\int
\let\tru@sum=\sum
\def\int{\mathop{\textstyle\tru@int}\limits}
\def\sum{\mathop{\textstyle\tru@sum}\limits}
\def\@#1{{\mathbf{#1}}}

\catcode`\@ 12

\let\^=\hat
\let\~=\tilde
\let\_=\bar

\newcommand{\bs}{\begin{subequations}}
\newcommand{\es}{\end{subequations}}

\def\XXint#1#2#3{{\setbox0=\hbox{$#1{#2#3}{\int}$}
     \vcenter{\hbox{$#2#3$}}\kern-.75\wd0}}

\def\@#1{{\mathbf{#1}}}

\newcommand{\CC}{\mathcal{C}}

\newcommand{\Z}{\mathbb{Z}}

 \def\RR{\mathbb R}

\let\^=\hat
\let\<=\langle
\let\>=\rangle

\newtheorem{theorem}{Theorem}

\newtheorem{proposition}[theorem]{Proposition}

\newtheorem{remark}[theorem]{Remark}

\title{Novel periodic solutions and rogue waves of the defocusing scalar and coupled Ablowitz-Ladik systems on a nonzero background}
\author{Francesco Coppini$^{1,2}$\; and Barbara Prinari$^{1,3,*}$}
\date{
    \small{$^{1}$ Department of Mathematics, State University of New York, Buffalo, NY 14260, USA\\
    $^2$ {Department of Physics, Università di Roma ``La Sapienza'', I-00185 Roma, Italy} \\
    $^4$ Department of Mathematics, University of Ioannina, Ioannina 45110, Epirus, Greece}\\
    $^*$ {Corresponding author: bprinari@buffalo.edu}
}

\begin{document}

\maketitle

\begin{abstract}
In this paper we apply Hirota’s bilinear method to the scalar and coupled Ablowitz-Ladik systems in the defocusing dispersion regime under the assumption of a background amplitude $0<\rho<1$. 
We first establish, in the scalar case, the correspondence between the Hirota's parameters and the spectral parameters of the inverse scattering transform. Then we show that when the Hirota parameter associated to the discrete eigenvalue is chosen outside the range corresponding to a discrete dark soliton, novel solutions of the Ablowitz-Ladik system emerge. In general, these solutions are singular, but there exists a class of time-periodic solutions for which it is possible to choose the soliton parameters so that the breathers remain regular on the lattice for all times. We also discuss the interactions between a dark soliton and a regular breather, and between two regular breathers. For the coupled Ablowitz-Ladik system, by including in the background discrete, counter-propagating plane waves, we use Hirota's method to derive novel Akhmediev-type (i.e., space-periodic) discrete breathers which are regular for all times. Finally, taking the limit of the discrete Akhmediev breathers as the period approaches infinity (i.e., as the wavenumber approaches zero) we obtain novel rogue wave solutions of the coupled Ablowitz-Ladik system.

\end{abstract}

\section{Introduction}

The study of nonlinear wave propagation in discrete media has attracted considerable attention over the past several decades, motivated both by fundamental mathematical questions and by a broad range of physical applications. Besides begin crucial for numerical schemes, discretizations arise naturally in applications such as optical waveguide arrays and lattice dynamical systems, biological systems, etc.
Among the most important integrable lattice models is the discrete nonlinear Schr\"odinger equation introduced by Ablowitz and Ladik \cite{AL1,AL2}, and commonly known as the Ablowitz–Ladik (AL) system: 
\begin{equation}
\label{ALv2}
 i\frac{dq_n}{dt}=q_{n+1}-2q_n+q_{n-1}-\sigma |q_n|^2(q_{n+1}+q_{n-1}) \qquad \sigma=\mp 1,
% (q_{n+1}+q_{n-1})(1-|q_n|^2)-nQ_n,
\end{equation}
where $q_n(t)$ is a complex function of $n\in \mathbb{N}$ and $t\in \Real$, and $\sigma=\mp 1$ distinguishes between the focusing and defocusing dispersion regimes, respectively.
This model constitutes an integrable discretization of the nonlinear Schr\"odinger (NLS) equation \cite{ZS72,ZS73} and preserves many of the remarkable properties of its continuous counterpart, including multisoliton solutions (both in the focusing and defocusing regimes), a Lax pair and associated Inverse Scattering Transform (IST), infinitely many conservation laws, etc. Because of these features, the AL system has become a prototypical framework for investigating nonlinear coherent structures in discrete settings.

While the rapidly decaying bright-soliton solutions of the focusing AL equation have been extensively studied since the mid 70's,  discrete analogs of the Tajiri-Watanabe \cite{TW98}, Kuznetsov-Ma \cite{K77,Ma79}, Akhmediev \cite{Akhmediev1985Generation,Akhmediev1986Exact},
and Peregrine \cite{Peregrine83} breathers which arise when a nonzero background is considered were only discovered in the last fifteen years \cite{AASC2010,Ank13,Ankiewicz2014,OY14,MRC26}, and the development of the IST on a nonzero background for the focusing AL system \cite{PV16}, together with the spectral characterization of the breathers \cite{P16}, are also much more recent.

The defocusing AL equation presents an equally rich phenomenology when a nonzero background is considered. 
In particular, it has long been known that when the background amplitude $\rho$ is less than 1, the defocusing AL system supports discrete dark solitons, and these solutions have been obtained both via the IST \cite{KV92,ABP07}, and via direct methods \cite{Ohta00,MO06}.  Although initially the constraint of a small background $0<\rho<1$ might have appeared merely as a technical requirement, it has now become apparent that the properties and the solutions of the defocusing AL equation crucially depend on whether the background amplitude is below or above 1. 
Indeed, Ohta \& Yang showed in \cite{OY14} that the defocusing AL system is modulationally stable in the small norm regime $0<\rho<1$, but it becomes unstable for $\rho>1$, and it exhibits discrete rogue wave solutions, some of which are regular for all times (see also \cite{CS24} for a study of the modulational instability of the discrete Akhmediev breathers in both focusing and defocusing regimes). The IST of the defocusing AL equation with a large background $\rho>1$ was developed in \cite{OP19} and, as a byproduct of the IST, novel discrete solutions were obtained which mimic the aforementioned discrete breathers of the focusing AL equation. Although the discrete breathers of the defocusing AL model with large background generically become singular in finite time, the discrete Kuznetsov-Ma (KM) breathers are special, in that, similar to the limiting case of the discrete Peregrine solutions found in \cite{OY14}, the breather parameters can be chosen so that the solution is regular on the lattice for all times \cite{WM25}. A novel KM-type breather solution which is regular on the lattice under the same conditions was also obtained in \cite{WM25}, and Darboux transformations were used to derive double KM breathers which are also regular on the lattice for suitable choices of the breathers' parameters.

Several complementary analytical approaches have been developed and used to construct exact solutions of integrable lattice equations. Among these, Hirota’s bilinear method occupies a distinguished role due to its direct and constructive nature. By rewriting the nonlinear evolution equation in bilinear form through suitable dependent-variable transformations, Hirota’s method enables the systematic derivation of exact solutions, including solitons, breathers, and rational solutions (rogue waves). For the AL system and related discrete integrable models, bilinear techniques have yielded a variety of explicit solutions and have provided important insight into the algebraic structure underlying nonlinear wave interactions. Nevertheless, the connection between the parameters appearing in Hirota’s formalism and the spectral parameters arising from the IST is not always completely evident, and this is the case for the defocusing discrete setting with nonzero background. Establishing such a correspondence is important both mathematically and physically, since it clarifies which bilinear solutions correspond to genuine discrete eigenvalues of the scattering problem and which instead may give rise to qualitatively different nonlinear states.

In the present work we investigate the scalar and coupled defocusing AL systems [the latter will also be referred later on as the integrable discrete Manakov (IDM) system] under the assumption of a small nonzero background. The analysis begins with the scalar AL equation, for which we derive an explicit correspondence between the parameters entering Hirota’s bilinear representation and the spectral parameters associated with the IST formulation. This mapping allows us to identify precisely the parameter regimes corresponding to discrete dark solitons, and it reveals that when the spectral parameter lies outside the range associated with discrete eigenvalues, new classes of nonlinear excitations emerge. In general, these solutions are singular, but for special parameter choices we obtain discrete breathers that are periodic in time and homoclinic in space which remain regular on the lattice for all times. These structures constitute discrete analogues of breather states propagating on a background amplitude greater than 1, and exhibit dynamical features absent in the standard dark-soliton sector.

We then consider the coupled AL system introduced as a vector generalization
of the AL equations  \cite{GI81a,GI81b,GI82,TUW99,Tsuchida00,AOT99}, and an integrable spatial discretization of the continuous Manakov system \cite{M74}:
\begin{equation}
\label{e:IDM}
\begin{split}
&  i\partial_tq^{(1)}_n+(q^{(1)}_{n+1}+q^{(1)}_{n-1})\left[1+\sigma(|q^{(1)}_n|^2+|q^{(2)}_n|^2)\right]=0,\\[2mm]
&  i\partial_tq^{(2)}_n+(q^{(2)}_{n+1}+q^{(2)}_{n-1})\left[1+\sigma(|q^{(1)}_n|^2+|q^{(2)}_n|^2)\right]=0,\\
    \end{split} \qquad \sigma=\mp 1
\end{equation}
where $\@q_n(t)=\left(q_n^{(1)}(t),q_n^{(2)}(t)\right)^T$ is a complex vector function of $n\in \mathbb{N}$ and $t\in \Real$, which we refer to as the integrable discrete Manakov (IDM) system. For the focusing ($\sigma=-1$) IDM system, the IST framework developed in \cite{APT04,APT04b,APT06} showed that different types of discrete bright solitons and breathers (fundamental solitons, fundamental breathers and composite breathers) exist, as the nature of the solution crucially depends on the rank of the associated norming constants (see also \cite{COP25} for a characterization of the soliton-breather interactions). On the other hand, while the generalization of the IST for the IDM system to a nonzero background is currently an open problem, solutions of multi-component defocusing AL systems have been obtained via direct methods in \cite{Ohta00,Ohta09,FO17}. In particular, \cite{FO17} confirmed that, like its continuous Manakov counterpart, the defocusing IDM system with nonzero background admits both discrete dark-dark solitons and discrete dark-bright solitons, with interaction properties that are similar to the ones in the continuous case \cite{DKPV13,PVB15}. Now, for the continuous defocusing Manakov system more general dark-dark and dark-bright solutions superimposed to a vector plane-wave background were constructed using the Darboux dressing method in \cite{DL07,DL09}, and it was subsequently shown that the plane-wave background induces in the system a base-band modulational instability which allows for the existence of vector rogues waves  \cite{Baronio14}. 

By introducing counter-propagating discrete plane-wave backgrounds in the Hirota's solutions, we derive novel Akhmediev-type discrete breathers for the defocusing IDM system. These solutions are periodic in space and homoclinic in time, and, unlike the scalar periodic states studied in \cite{OP19,CS24}, there exist parameter regimes for which the solutions remain regular on the lattice for all times when, as assumed here, the norm of the background is less than 1. Finally, taking the limit of the discrete Akhmediev breathers as the wavenumber approaches zero (i.e., as the period approaches infinity), we obtain rogue wave solutions of the coupled AL system.  To the best of our knowledge, these classes of exact discrete breathers and rogue waves have not previously been reported.

The results presented here contribute to the broader program of understanding nonlinear coherent structures in integrable discrete media with nonzero boundary conditions. In particular, they highlight how the interplay between bilinear methods and inverse scattering theory can uncover new classes of exact solutions beyond the traditional soliton regime. They also provide further evidence that the defocusing AL hierarchy possesses a substantially richer solution space than previously recognized, especially in the vector setting where background interactions generate novel breather dynamics.

The plan of the paper is the following. In Sec.~2 we review the spectral properties of the discrete dark solitons of the defocusing AL lattice on a background amplitude $0<\rho<1$, and relate them to the corresponding solutions which we obtain via Hirota's method. The latter are more general dark solitons, as they are superimposed on a (discrete) plane-wave background. Then we construct novel breather solutions of the AL equation which correspond to different parameter choices than the dark soliton case. In general, these solutions are singular. However, when reduced to the case of a constant background, we show that there exist choices of the breather parameters for which the breathers which are periodic in time and homoclinic in space (i.e., discrete KM breathers) remain regular for all times on the lattice.  We also discuss the interactions between a dark soliton and a regular breather and between two regular breathers. In Sec.~3 we use Hirota's method to obtain discrete dark-bright solitons on a general 2-component discrete plane wave background for the IDM system. In the special case of counter-propagating plane waves, we obtain two novel classes of solutions which are periodic in space and homoclinc in time, i.e., discrete Akhmediev-type breathers, and, in the limit as the wavenumber approaches 0 (i.e., the limit of infinite period) coupled rational solutions are obtained. Finally, Sec.~4 offers some concluding remarks.

%The Hirota method belongs to the class of direct methods for constructing exact solutions of integrable models (although it has also been generalized to some cases of non-integrable equations). It is based on the search for solutions written in rational form; by introducing two arbitrary functions for the numerator and the denominator, the original equation is then suitably split into two parts so as to preserve the integrability of the theory and the absence of resonant terms. In this work we will use the Hirota method for the construction of solutions of the integrable discrete Manakov model in Section 2, but previously we will apply the method widely known in the literature for constructing solutions in the scalar case (Section 1).

\section{Defocusing AL equation}

%\subsection{Dark solitons of the defocusing AL with nonzero background $0<\rho<1$}
We consider here the defocusing AL in the form:
\begin{equation}
\label{e:AL}
 i\frac{dq_n}{dt}%=q_{n+1}-2q_n+q_{n-1}-|q_n|^2(q_{n+1}+q_{n-1})\equiv    
 =(q_{n+1}+q_{n-1})(1-|q_n|^2) %-2q_n
\end{equation}
with boundary conditions:
\begin{equation}
q_n(t) \sim \rho\, e^{i\theta_\pm +2i\rho^2 t} \qquad n\to \pm \infty, \qquad \text{with } 0<\rho<1, \quad \theta_\pm \in \Real.    
\end{equation}
For comparison purposes, note that the equation is often also written in the form \eqref{ALv2}
%\begin{equation}
%\label{ALv2}
%i\frac{du_n}{dt}=(1-|u_n|^2)(u_{n+1}+u_{n-1})-2u_n,
%\end{equation}
(cf., e.g., \cite{ABP07}), since it is the form that naturally arises when one considers a second order difference as spatial discretization for the second derivative in the NLS equation, but the additional linear term $-2q_n$ can be easily restored by a simple gauge transformation $q_n\mapsto q_n e^{2i t}$.

\subsection{Dark solitons of the defocusing AL with small background via IST}
We review below the details of the construction of the dark soliton from the IST as developed in \cite{ABP07}, with the goal of then establishing the precise correspondence between the soliton spectral parameters and the parameters that appear in Hirota's method. The derivation follows closely the one in \cite{ABP07}, with minor, self-explanatory changes in notation. We also recall that in \cite{ABP07} the defocusing AL equation was written in the form \eqref{ALv2}, so the solution will have ad additional (inessential) phase factor $e^{2it}$ compared to the ones obtained in this work. Let $z$ denote the spectral parameter in the standard AL Lax pair, define:
\begin{equation}
    r=\sqrt{1-\rho^2}, \qquad 0<r<1,
\end{equation}
and the variables:
\begin{gather}
\xi=\xi(z)=\frac{z+1/z}{2r}, \qquad \lambda=\lambda(z)=\xi\pm \sqrt{\xi^2-1} \qquad \text{s.t. } 
1/\lambda(z)=\xi\mp \sqrt{\xi^2-1}.
\end{gather}
The discrete analog of the plane wave has $\lambda^{2n},\lambda^{-2n}$; and $\lambda(z)$ has branch points at the 4 points:
$$
\pm z_o=\pm(r+i\rho), \qquad \pm z_o^*=\pm (r-i\rho).
$$
All 4 branch points are (symmetrically) located on the circle of radius 1 (see Fig.~2 in \cite{ABP07}); the upper and lower arcs of the circle are chosen as the branch cuts, and the discrete eigenvalues are shown to be located on the left and right arcs; specifically:
\begin{subequations}
\label{conds_zk}
\begin{equation}
\label{e:zbk}
 \bar{z}_k=e^{i\alpha_k}\equiv \cos (\alpha_k)+i\sin(\alpha_k), \qquad  -\frac{\rho}{r}<\tan \alpha_k < \frac{\rho}{r}  
\end{equation}
with $r<\Re(\bar{z}_k)\le 1$, or symmetrically located on the left arc, with  $-1\le \Re(\bar{z}_k)<-r$.

Each discrete eigenvalue is such that:
\begin{gather}
|\Re \bar{z}_k|>r, \qquad |\bar{z}_k|=1, \\
\xi(\bar{z}_k)=\frac{\Re \bar{z}_k}{r}>1, \\
\lambda(\bar{z}_k)=\xi(\bar{z}_k)-\sqrt{(\xi(\bar{z}_k))^2-1}\equiv \frac{1}{r}\left[ \Re \bar{z}_k-\sqrt{(\Re \bar{z}_k)^2-r^2}\right]>0, \label{e:lambda(z)} \\
\lambda(\bar{z}_k)-1/\lambda(\bar{z}_k)=-2\sqrt{(\xi(\bar{\zeta}_k))^2-1}.
\end{gather}
\end{subequations}
The dispersion relation is given by:
\begin{equation}
\mu(z)=r(z-1/z)(\lambda-1/\lambda),    
\end{equation}
and the norming constant $\bar{C}_k$ associated to the egienvalue $\bar{z}_k$ evolves in time according to:
\begin{equation}
\bar{C}_k(t)=\bar{C}_k(0)\exp \left[-2i\rho^2 t-i\mu(\bar{z}_k)t \right],    
\end{equation}
from which one obtains:
\begin{equation}
|\bar{C}_k(t)|=|\bar{C}_k(0)|\exp \left[ \Im \mu(\bar{z}_k) t\right].   
\end{equation}
Using the relations above, one can obtain
\begin{equation}
\mu(\bar{z}_k)=-4i \Im \bar{z}_k\sqrt{(\Re \bar{z}_k)^2-r^2}    
\end{equation}
which is purely imaginary, and in turn:
\begin{equation}
|\bar{C}_k(t)|=|\bar{C}_1(0)|\exp \left[ -4t \Im \bar{z}_k \sqrt{(\Re \bar{z}_k)^2-r^2}   \right]. 
\end{equation}
We now consider the explicit formula for a single discrete dark soliton, Eq.~(6.8) in \cite{ABP07}. It should be noted that the coefficients $a$ and $d$ in (6.8) are not independent, and we can rewrite the solution as:
\begin{equation}
Q_n^*(t)=\rho e^{-i\theta_+ -2i\rho^2 t}\left[ 1+ \frac{r\rho^{-1}e^{i\phi_1}(1-|\bar{\zeta}_1|^2)d \, \lambda^{2n}}{1+|\bar{\zeta}_1|d\, \lambda^{2n}} \right],    
\end{equation}
where
\begin{equation}
d=|\bar{C}_1(t)|(|\bar{\zeta}_1|-|\bar{\zeta}_1|^{-1}).    
\end{equation}
Note that $\lambda^{2n}$ stands for $(\lambda^2(\bar{z}_1))^n$ where $\lambda(\bar{z}_1)$ is given in \eqref{e:lambda(z)} (or $(\lambda^2(\bar{\zeta}_1))^n$ where $\lambda^2(\bar{z}_1)$ is given in terms of the uniformization variable in \eqref{e:z2lambda2zlambda}).
The additional parameters one needs to specify are $\bar{\zeta}_1\equiv 1/\zeta_1^*$ and $\phi_1=\arg (\bar{\zeta}_1-r)$, which can all be expressed in terms of the uniformization variable. 
In order to express everything in terms of $\bar{z}_1$, we recall the following. The uniformization variable is defined as:
\begin{subequations}
\begin{equation}
\zeta=\frac{\lambda(z)}{z},    
\end{equation}
and it is such that
\begin{equation}
\label{e:z2lambda2zlambda}
z^2=\frac{\zeta-r}{\zeta(r\zeta-1)}, \qquad \lambda^2=\frac{\zeta(\zeta-r)}{r\zeta-1}, \qquad z\lambda-\frac{\zeta-r}{r\zeta-1}.    
\end{equation}
\end{subequations}
In terms of $\zeta$, the discrete eigenvalues outside $|\zeta|=1$ are $\zeta_k$ such that $|\zeta_k-1/r|=\rho/r$. As usual
$$\bar{\zeta}_k=1/\zeta_k^*$$ and we want $|\bar{\zeta}_k|<1$. Then, according to Eq.~(2.76) and Fig.~3 in \cite{ABP07}, this corresponds to:
\begin{equation}
\label{psi_k}
\bar{\zeta}_k=(1+\rho e^{i\psi_k})/r, \qquad |\pi -\psi_k|<\arctan (r/\rho).    
\end{equation}
%{\color{red}[this last conditions should be double-checked]}. 
We also note that there is a constraint on the phase of the norming constant coming from Eq.~(3.7) in \cite{ABP07}, which specifies that
\begin{equation}
\label{phase_C}
\arg \bar{C}_k(0)=-\theta_++\psi_k    \qquad \text{or } \quad \arg \bar{C}_k(0)=\pi-\theta_++\psi_k
\end{equation}
where $\psi_k$ is the angle defined in \eqref{psi_k}.

For our purposes, the identification of the parameters then proceeds as follows. We choose $\bar{z}_1$ as in \eqref{e:zbk}, then first obtain $\zeta_1$ as:
\begin{equation}
\zeta_1=\frac{\lambda(\bar{z}_1)}{\bar{z}_1}=\frac{\Re \bar{z}_1-\sqrt{(\Re \bar{z}_1)^2-r^2}}{r\bar{z}_1},    
\end{equation}
and then 
\begin{equation}
  \bar{\zeta}_1=\frac{1}{\zeta_1^*}\equiv \frac{r\bar{z}_k^*}{\Re \bar{z}_k-\sqrt{(\Re \bar{z}_k)^2-r^2}}.  
\end{equation} In particular, this implies that
\begin{equation}
|\bar{\zeta}_1|=|\zeta_1|^{-1}=\frac{|\bar{z}_1|}{|\lambda(\bar{\zeta}_1)|}\equiv |\lambda(\bar{\zeta}_1)|^{-1}   \equiv \frac{r}{\Re \bar{z}_1-\sqrt{(\Re \bar{z}_1)^2-r^2}}, 
\end{equation}
while the angle $\psi_1$ is determined by \eqref{psi_k} as:
\begin{equation}
\psi_1=\arg \frac{r\bar{\zeta}_1-1}{\rho}  \qquad \text{with } \quad |\pi -\psi_1|<\arctan (r/\rho).  
\end{equation}
The final parameter $\phi_1$ is determined as 
$$
\phi_1=\arg (\bar{\zeta}_1-r),
$$
which, according to (A.22) in \cite{ABP07}, is given in terms of $\psi_1$ as follows:
\begin{equation}
\tan \phi_1=\frac{\sin \psi_1}{\rho+\cos \psi_1}.    
\end{equation}
Finally, Eq.~(A.20) in \cite{ABP07} also shows that the correct choice for the phase of the norming constant in order to get a regular solution is the second in \eqref{phase_C}, i.e.:
\begin{equation}
\arg \bar{C}_1(0)=\pi -\theta_++\psi_1.     
\end{equation}
\subsection{Hirota's method for the defocusing AL system with small background}
We now deal with the construction of explicit solutions of the defocusing AL equation with a small background via Hirota's method. We look for a solution in rational form:
\begin{equation}
    q_n=\rho\,e^{i(k_0 n+\omega_0 t)}\;\frac{g_n}{f_n},
\end{equation}
with $f_n\in\RR$ and $g_n\in\mathbb{C}$ and the dispersion relation 
\begin{equation}
\omega_0(k_0)\equiv\omega_0=2(1-\rho^2)\cos  k_0.
\end{equation}
Substituting this ansatz into \eqref{e:AL} we obtain:
\begin{equation}
\begin{split}
i\frac{(\partial_t g_n)f_n-(\partial_t f_n)g_n}{f_n^2} -2&(1-\rho^2)\cos k_0   \frac{g_n}{f_n}+\frac{e^{ik_0} g_{n+1}f_{n-1}+e^{-ik_0} g_{n-1}f_{n+1} }{f_{n-1}f_{n+1}}+\\[2mm]
-&\rho^2\frac{e^{ik_0} g_{n+1}f_{n-1}+e^{-ik_0} g_{n-1}f_{n+1} }{f_{n-1}f_n^2f_{n+1}}|g_n|^2=0.
\end{split}
\end{equation}\\
We then add and subtract the term:
\begin{equation}
    (1-\rho^2)\frac{e^{ik_0} g_{n+1}f_{n-1}+e^{-ik_0} g_{n-1}f_{n+1} }{f_n^2},
\end{equation}
and rearrange the above expression as:
{\small\begin{equation}
\begin{split}
 \frac{1}{f_n^2}& \left[i(\partial_t g_n)f_n-i(\partial_t f_n)g_n +(1-\rho^2)(e^{ik_0} g_{n+1}f_{n-1}+e^{-ik_0} g_{n-1}f_{n+1} -2\cos k_0  \; g_nf_n) \right]+\\[2mm]
  +&\frac{(e^{ik_0} g_{n+1}f_{n-1}+e^{-ik_0} g_{n-1}f_{n+1})}{f_{n-1}f_n^2f_{n+1}}\left[f_n^2-\rho^2|g_n|^2-(1-\rho^2)f_{n-1}f_{n+1}\right].
  \end{split}
\end{equation}}
The correct way to separate the above expression is to isolate the two terms inside the square brackets:\\
{\small\begin{equation}
\begin{split}
  &  H_1[g_n,f_n]:=i(\partial_t g_n)f_n-i(\partial_t f_n)g_n +(1-\,\rho^2)(e^{ik_0} g_{n+1}f_{n-1}+e^{-ik_0} g_{n-1}f_{n+1} -2\cos k_0  \; g_nf_n), \\[2mm]
   & H_2[g_n,f_n]:=f_n^2-\rho^2|g_n|^2-(1-\rho^2)f_{n-1}f_{n+1}.
   \end{split}
\end{equation}}\\
We can rewrite the two equations above in terms of the continuous Hirota derivative operator $ D_t$:
\begin{equation}
    D_t(a\cdot b)=\left(\partial_t-\partial_{t'}\right)a(t)b(t')|_{t'=t}
\end{equation}\\
and the discrete derivative\\
\begin{equation}
 e^{D_n}(a\cdot b)\equiv  e^{\partial_n-\partial_{m}}\,a_nb_{m}\,|_{m=n}=a_{n+1}b_{n-1}.
\end{equation}
Specifically, one has:
\begin{equation}
\begin{split}
  &  H_1[g_n,f_n]:=iD_t(g_n\cdot f_n)+4(1-\rho^2)\sinh\left(\frac{D_n}{2}\right)\sinh\left(\frac{D_n}{2}+ik_0\right)( g_n\cdot f_n),\\[2mm]
   & H_2[g_n,f_n]:=-\rho^2\left(|g_n|^2-f_n^2\right)-(1-\rho^2)\sinh\left(\frac{D_n}{2}\right)^2(f_{n}\cdot f_{n}).
   \end{split}
\end{equation}
\begin{equation}\label{Hirota_sol_def}
    q_n=\rho\,e^{i(k_0 n+2(1-\,\rho^2)\cos  k_0 t)}\;\frac{g_n}{f_n}.
\end{equation}\\
The two Hirota operators are specified by the following $H_1$ and $H_2$:
\begin{equation}\label{Hirota_defocusing}
\begin{split}
  &  H_1[g_n,f_n]:=i(\partial_t g_n)f_n-i(\partial_t f_n)g_n +(1-\rho^2)(e^{ik_0} g_{n+1}f_{n-1}+e^{-ik_0} g_{n-1}f_{n+1} -2\cos k_0  \; g_nf_n),\\[2mm]
   & H_2[g_n,f_n]:=f_n^2-\rho^2|g_n|^2-(1-\rho^2)f_{n-1}f_{n+1}.
   \end{split}
\end{equation}
As usual, we start by writing for the functions $f_n$ and $g_n$ expansions with respect to a small parameter $\epsilon\ll 1$:
\begin{gather}
       f_n=f^{(0)}_n+\epsilon f^{(1)}_n+\epsilon^2 f^{(2)}_n, \qquad 
     g_n=g^{(0)}_n+\epsilon g^{(1)}_n+\epsilon^2 g^{(2)}_n.
\end{gather}
Since we are interested in studying solutions on a nonzero background $\rho$, we anchor the perturbative expansion with:
\begin{equation}
    f^{(0)}_n=g^{(0)}_n=1.
\end{equation}
Once the perturbative series is substituted into the Hirota operators, we obtain the iterative expansion of the successive orders of the Taylor series in $\epsilon$. At leading order O($\epsilon$), the linearized operators $L_1$ and $L_2$ appear:
\begin{subequations}
\label{Linear_Hirota_def}
\begin{align}
 & L_1[g^{(1)}_n,f^{(1)}_n]\equiv \\
&  i\partial_t(g^{(1)}_n-f^{(1)}_n)  +(1-\rho^2)\left( e^{ik_0}g^{(1)}_{n+1}+e^{-ik_0}g^{(1)}_{n-1}-2g^{(1)}_{n}+ e^{ik_0}f^{(1)}_{n-1}+e^{ik_0}f^{(1)}_{n+1}-2f^{(1)}_{n}\right)   =0,  \nonumber \\
&  L_2[g^{(1)}_n,f^{(1)}_n]\equiv -\rho^2\left(g^{(1)}_n+(g^{(1)}_n)^*-2f^{(1)}_n\right)-(1-\rho^2)\left(f^{(1)}_{n+1}+f^{(1)}_{n-1}-2f^{(1)}_{n}\right)=0.
\end{align}
\end{subequations}
As usual, at higher orders, the same linearized operators recur, and the functions computed at previous orders appear as forcing terms:
\begin{subequations}
\begin{align}
  &L_1[g^{(2)}_n,f^{(2)}_n]= \\
  &=-\left[i\left(f^{(1)}_n\,\partial_t g^{(1)}_n-g^{(1)}_n\,\partial_t f^{(1)}_n\right)+(1-\rho^2)\left(e^{i k_0}f^{(1)}_{n-1}\,g^{(1)}_{n+1}+e^{-i k_0}f^{(1)}_{n+1}\,g^{(1)}_{n-1}-2\cos k_0 f^{(1)}_{n}\,g^{(1)}_{n}\right)\right], \nonumber \\
  &L_2[g^{(2)}_n,f^{(2)}_n]=(1-\rho^2)f^{(1)}_{n-1}\,f^{(1)}_{n+1}+\rho^2|g^{(1)}_{n}|^2-\left(f^{(1)}_{n}\right)^2,
\end{align}
\end{subequations}
and so on for all subsequent orders. If we are interested in a solution localized on the background, we look for $f^{(1)}_n,g^{(1)}_n$ in the form of real exponentials:
\begin{equation}\label{linear_Hirota_def}
    g^{(1)}_n=\alpha \,e^{k n-\sigma t}, \qquad f^{(1)}_n=e^{k n-\sigma t}.
\end{equation}
Substituting this anzatz into \eqref{Linear_Hirota_def} we obtain the following constraints:
\begin{equation}
\begin{split}
    &L_1[g_n^{(1)},f_n^{(1)}]=e^{kn-\sigma t}\bigg[ \alpha\left(4(1-\rho^2)\sinh^2\left(\frac{k}{2}\right)\cos k_0-i\left(\sigma-2(1-\rho^2)\sinh k \sin k_0\right)\right)+ \\
    &\   \ \ \ \  \  \ \ \ \ \ \ \  \  \ \ \ \  \  \  \  \  \  \   \   \  \ \  \ \ +\left(4(1-\rho^2)\sinh^2\left(\frac{k}{2}\right)\cos k_0+i\left(\sigma-2(1-\rho^2)\sinh k \sin k_0\right)\right) \bigg] =0\\[2mm]
   & L_2[g_n^{(1)},f_n^{(1)}]=2\rho^2e^{kn-\sigma t}\left[ 1-\Re \alpha-2\frac{(1-\rho^2)}{\rho^2}\sinh^2 \left(\frac{k}{2}\right) \right]=0.
  \end{split}  
\end{equation}
This yields a system of one complex equation and one real equation for a complex variable $\alpha$ and a real variable $\sigma$, whose solution is given by:
\begin{equation}\label{alpha_sigma}
    \begin{split}
    &\alpha=1-2\delta^2\mp 2i\delta\sqrt{1-\delta^2},\\
    &\sigma=\pm 4\rho^2\delta\sqrt{1-\delta^2}\cos k_0+2(1-\rho^2)\sinh k\sin k_0,
    \end{split}
\end{equation}
where:
\begin{equation}
    \delta=\sqrt{\frac{1-\rho^2}{\rho^2}}\sinh\left(\frac{k}{2}\right).
\end{equation}\\
Since we derived the solution assuming $\sigma\in\Real$, it is clear that one needs $\rho<1$ and $\delta<1$. 
In the case where $\delta<1$,  for all subsequent orders $O(\epsilon^j)$ for $j\geq 2$ there is no forcing term present:
\begin{equation}
L_1[g_n^{(j)},f_n^{(j)}]=L_2[g_n^{(j)},f_n^{(j)}]=0, \qquad j\ge 2,
\end{equation}
and therefore the series terminates, and  $g_n^{(j)}=f_n^{(j)}=0$ for $j\geq 2$.
In the following, we consider the two cases $\delta>1$ and $\delta<1$ separately, since in the latter case the ansatz for \eqref{Linear_Hirota_def} will be slightly modified.
\subsubsection{Solutions with $\delta<1$: dark solitons}
\begin{figure}[t!!]
\centering
\includegraphics[width=7cm,height=5.9cm]{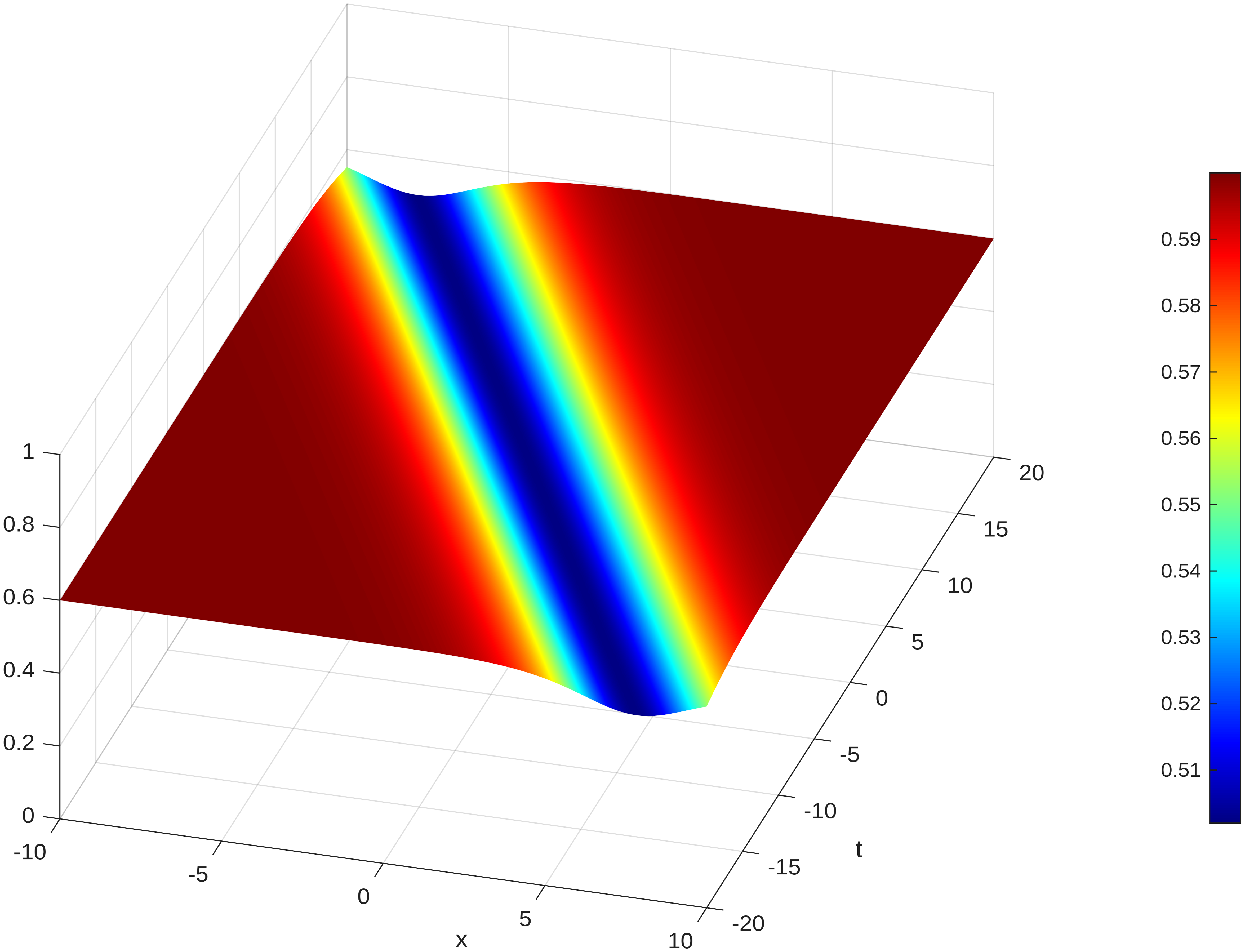}
\caption{Plot of the dark soliton \eqref{dark1} with parameters $k_0=\frac{2\pi}{8.2}$, $\rho=0.6$, $\xi_0=0$ and $k=0.8$, therefore the solution satisfies the existence condition $\delta=0.55<1$ .}
\label{f1}
\end{figure}

If  $\delta<1$ we can introduce the new angle $\theta$ defined by:
\begin{equation}\label{theta}
   \cos\theta\equiv \delta= \frac{  \sqrt{1-\rho^2}}{\rho}\sinh\left(\frac{k}{2}\right),
\end{equation}
in terms of which:
\begin{equation}
\begin{split}
  &\sigma_{\pm}=\pm 2\rho^2\sin 2\theta\cos k_0+2(1-\rho^2)\sinh k\sin k_0,\\
  &\alpha_{\pm}=-e^{\pm 2i\theta}.
  \end{split}
\end{equation}
In this case, the solution \eqref{Hirota_sol_def} reads:
\begin{equation}
     q_n^{\pm}(t)=\rho\,e^{i(k_0 n+\omega_0 t)}\;\frac{1-\epsilon \,e^{\pm 2i\theta+k n-\sigma_{\pm} t}}{1+\epsilon \,e^{k n-\sigma_{\pm} t}}.
\end{equation}
We can write the solution in a compact way introducing the shifting parameter $\xi_0=-\frac{1}{2}\log\epsilon$:
\begin{equation}\label{dark1}
    q_n^{\pm}(t) %=\rho\, e^{i(k_0 n+\omega_0 t\pm\theta)}\frac{\sinh\left(\frac{k n -\sigma_{\pm} t-\xi_0}{2}\pm i\theta\right)}{\cosh\left(\frac{k n -\sigma_{\pm} t-\xi_0}{2}\right)}
    =\rho\, e^{i(k_0 n+\omega_0 t)}\left(\pm i\sin\theta+\cos\theta\tanh\left(\frac{k n -\sigma_{\pm} t-\xi_0}{2}\right)\right),
\end{equation}
where   $\omega_0\equiv\omega_0(k_0)=2(1-\rho^2)\cos  k_0$ is the plain wave dispersion relation. This solution, plotted in Fig.~\ref{f1} represents a dark soliton superimposed on a plane-wave background.
To connect these results with those obtained using the IST in Sec.~2.1, it is sufficient to note that:
\begin{equation}
    \begin{split}
       & \lambda=e^{\frac{k}{2}},\\
       &\xi=\cosh\frac{k}{2},\\
       &\cos\alpha_1=\frac{1}{2}\left(z_1+\frac{1}{z_1}\right)=\sqrt{1-\rho^2}\cosh\frac{k}{2}
    \end{split}
\end{equation}
If $k $ and $\alpha_1$are required to be real parameters, the previous formula imposes the appropriate constraints on the two parameters; in particular, one obtains:
\begin{equation}
    \cos\alpha_1>\sqrt{1+\rho^2}, \qquad \cosh\frac{k}{2}<\frac{1}{\sqrt{1-\rho^2}}.
\end{equation}\\
It is easy to prove that the last condition is equivalent to $ \sinh\left(\frac{k}{2}\right)<\frac{\rho}{  \sqrt{1-\rho^2}}$, that was the basic assumption we introduced during the derivation of the dark soliton solution with the Hirota method.
\subsubsection{Solutions with $\delta>1$: KM breathers}
When $\delta>1$ , the ansatz for the solution must be slightly modified to account for the fact that the velocity $\sigma$ as defined in \eqref{alpha_sigma} is now a complex number, by taking:
\begin{equation}\label{linear_Hirota_def}
    g^{(1)}_n=\alpha \,e^{k n-\sigma_r t+i\sigma_i (t-t_0)}+\beta \,e^{k n-\sigma_r t-i\sigma_i (t-t_0)}, \qquad  f^{(1)}_n=e^{k n-\sigma_r t}\cos(\sigma_i (t-t_0)).
\end{equation}
In this case the action of the operators $L_1$ and $L_2$ give:
\begin{equation}
\begin{split}
    &L_1[g_n^{(1)},f_n^{(1)}]=e^{kn-(\sigma_r-i\sigma_i) t-i \sigma_i t_0}\bigg[ \alpha\left(\sigma_i-4\rho^2\delta^2\cos k_0+i(\sigma_r-2(1-\rho^2)\sinh k\sin k_0)\right)+ \\
    &\   \ \ \ \  \  \ \ \ \ \ \ \  \  \ \ \ \  \  \  \  \  \  \   \   \  \ \  \ \ +\left(-\sigma_i-4\rho^2\delta^2\cos k_0-i(\sigma_r-2(1-\rho^2)\sinh k\sin k_0)\right) \bigg] +\\[2mm]
    &+e^{kn-(\sigma_r+i\sigma_i) t+ i\sigma_i t_0}\bigg[ \beta\left(\sigma_i+4\rho^2\delta^2\cos k_0-i(\sigma_r-2(1-\rho^2)\sinh k\sin k_0)\right)+ \\
    &\   \ \ \ \  \  \ \ \ \ \ \ \  \  \ \ \ \  \  \  \  \  \  \   \   \  \ \  \ \ +\left(-\sigma_i+4\rho^2\delta^2\cos k_0+i(\sigma_r-2(1-\rho^2)\sinh k\sin k_0)\right) \bigg]=0, \\[2mm]
   & L_2[g_n^{(1)},f_n^{(1)}]=-\rho^2e^{kn-(\sigma_r+i\sigma_i) t+ i\sigma_i t_0}\left[  \alpha+\beta^*-2+4\delta^2 \right]+c.c.=0,
  \end{split}  
\end{equation}
where $c.c.$ denotes the complex conjugate first term. The above equations yield the following constraints on the Hirota parameters:
\begin{equation*}
    \begin{split}
        &\sigma_r=2(1-\rho^2)\sinh k\sin k_0, \qquad \sigma_i^\pm=\pm 4\rho^2\delta\sqrt{\delta^2-1}\cos k_0,\\
        &\alpha^\pm=1-2\delta^2\pm 2\delta\sqrt{\delta^2-1},\qquad \beta^\pm =1/\alpha^{\pm},
    \end{split}
\end{equation*}
with arbitrary $t_0\in\RR$. In this case, it is convenient to introduce a hyperbolic variable $\phi$:
\begin{equation}\label{phi}
  \delta\equiv\cosh\phi= \frac{  \sqrt{1-\rho^2}}{\rho}\sinh\left(\frac{|k|}{2}\right),
\end{equation}
which allows to simplify the above quantities 
\begin{equation}
\alpha^\pm=-e^{\pm 2\phi}, \qquad \beta^\pm =-e^{\mp 2\phi}, \qquad \sigma_i^\pm=\pm 2\rho^2\sinh(2\phi)\cos k_0. 
\end{equation}
By examining the second order $O(\epsilon)$ of the perturbative series, we then obtain:
\begin{equation}
L_1[g_n^{(2)},f_n^{(2)}]=-4\sinh(2\phi)e^{2(kn-\sigma_r t)}, \qquad 
L_2[g_n^{(2)},f_n^{(2)}]=8\rho^2\sinh^2(2\phi)e^{2(kn-\sigma_r t)},
\end{equation}
whose solution is given by:
\begin{equation}
    g_n^{(2)}=f_n^{(2)}=-a_{11}\,e^{2(k n -\sigma_r t)}, \qquad a_{11}=-\frac{\sinh^2 \phi}{\cosh^2\frac{k}{2}}.
\end{equation} 
From the third order onward, no forcing terms arise, so the perturbative series can be truncated. The final solution reads: 
\begin{equation}
     q_n^{\pm}(t)=\rho\,e^{i(k_0 n+\omega_0 t)}\;\frac{1-\epsilon\, e^{ k n-\sigma_r t} \left(e^{\pm (i\sigma_i\,(t-t_0)+ 2\phi)}+e^{\mp (i\sigma_i\,(t-t_0)+2\phi)}\right)+a_{11}\,\epsilon^2\,e^{2(k n -\sigma_r t)}}{1+2\, \epsilon\,  e^{ k n-\sigma_r t}\cos(\sigma_i\,(t-t_0))+a_{11}\epsilon^2 \,e^{2(k n -\sigma_r t)}}.
\end{equation}
It is easy to see that the different signs in the solution are irrelevant; therefore, without loss of generality, we proceed with the analysis by choosing the solution with the positive sign: $q_n(t)\equiv q_n^{+}(t)$ and $\sigma_i\equiv\sigma_i^{+}=2\rho^2\sinh(2\phi)\cos k_0$.
Introducing the shifting parameter: 
\begin{equation}
\label{e:center}
{\xi_0}=-\ln (\epsilon\sqrt{|a_{11}|}),
\end{equation}
we can write the solution in the compact form:
\begin{equation}\label{dark_singulare}
\begin{split}
    q_n(t)
    & =\rho\, e^{i(k_0 n+\omega_0 t)}\frac{\sinh\left(k n -\sigma_r t-\xi_0\right)+\frac{\cosh\left(\frac{k}{2}\right)}{\sinh\phi}\cos\left(\sigma_i (t-t_0)-2i\phi\right)}{\sinh\left(k n -\sigma_r t-\xi_0\right)-\frac{\cosh\left(\frac{k}{2}\right)}{\sinh\phi}\cos(\sigma_i (t-t_0))}=\\[3mm]
    =&\rho\, e^{i(k_0 n+\omega_0 t)}\left[1+2\frac{\cosh\left(\frac{k}{2}\right)}{\tanh\phi}\frac{\cos(\sigma_i (t-t_0))\cosh\phi+i\sin(\sigma_i (t-t_0))\sinh\phi}{\sinh\left(k n -\sigma_r t-\xi_0\right)-\frac{\cosh\frac{k}{2}}{\sinh\phi}\cos\left(\sigma_i(t-t_0)\right)}\right].
    \end{split}
\end{equation}
The above solution is similar to the regular breathers one obtains for the focusing AL on a nonzero background, and to the (singular) breathers that exist in the defocusing regime when the background $\rho>1$ \cite{OP19,WM25}.
Due to the presence of the hyperbolic sine in the denominator, the solution will in general develop singularities along a curve. However, similarly to what happens in the case of large background, there are special cases in which this singular curve lies within the lattice spacing. In particular, if the wave vector $k_0=0$ of the plane wave is chosen to be zero, one obtains solutions with vanishing group velocity $\sigma_r=0$:
\begin{equation}\label{KM}
    q_n(t)
    %& =\rho\, e^{2i(1-\rho^2)t}\frac{\sinh\left(k n -\sigma_r t-\xi_0\right)+g\cos\left(\sigma_i t-2i\phi+\zeta_0\righ&\textcolor{red}{
    =\rho\, e^{i(k_0 n+\omega_0 t)}\left[1+2\frac{\cosh\left(\frac{k}{2}\right)}{\tanh\phi}\frac{\cos(\sigma_i (t-t_0))\cosh\phi+i\sin(\sigma_i (t-t_0))\sinh\phi}{\sinh\left(k n-\xi_0\right)-\frac{\cosh\frac{k}{2}}{\sinh\phi}\cos\left(\sigma_i(t-t_0)\right)}\right],
\end{equation}
which are periodic in $t$ and homoclinic in $n$, and very similar to the KM breathers in the case of large background \cite{WM25}, the main difference being the hyperbolic sine instead of an hyperbolic cosine in the denominator. In both cases, the solutions are generically singular. In this case, the singular curve $n(t)$ is defined by the relation:
\begin{equation*}
    \sinh(k\, n-\xi_0)=\frac{\cosh\frac{k}{2}}{\sinh\phi}\cos(\sigma_i (t-t_0)).
\end{equation*}
or equivalently:
\begin{equation}\label{n_t_singular}
    n(t)=\frac{1}{k}\left[\xi_0+\textrm{arcsinh}\left(\frac{\cosh\frac{k}{2}}{\sinh\phi}\cos(\sigma_i (t-t_0))\right)\right].
\end{equation}
\begin{figure}[t!!]
\label{f2}
\centering
\includegraphics[width=5cm,height=5cm]{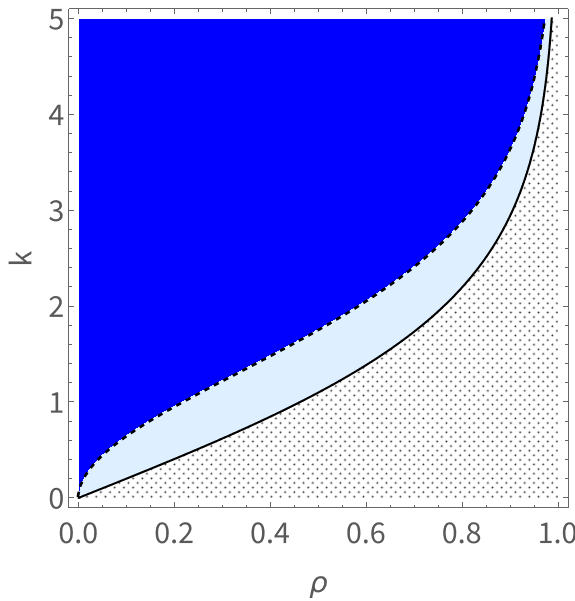}
 \caption{Plot of the curve $g(\rho,k)=1$ (black dashed line), and the curve $\delta=\frac{  \sqrt{1-\rho^2}}{\rho}\sinh\left(\frac{k}{2}\right)=1$ (black solid line). The dotted region represents the domain where the solution is regular. The light shaded region corresponds to singular solutions whose singular curve has a width larger than the lattice spacing. The dark region denotes singular solutions  whose singular curve has an amplitude less than the lattice spacing for which the function can still be “regularized”.}
\end{figure}
Similarly to what was done in \cite{WM25} in the case $\rho>1$, in order to obtain solutions that are regular on the lattice for all times, we require that the support of the singular curve be contained within the lattice spacing; as an example, we take the first positive integer $0<n(t)<1$, i.e., require:
\begin{equation}\label{reg_cond_1}
 g(\rho,k)\equiv   \frac{2}{k}\,\textrm{arcsinh}\left(\frac{\cosh\frac{k}{2}}{\sinh\phi}\right)=  \frac{2}{k}\,\textrm{arcsinh}\left(\frac{\cosh\frac{k}{2}}{\sqrt{\frac{1-\rho^2}{\rho^2}\cosh^2\frac{k}{2}-\frac{1}{\rho^2}}}\right)<1,
\end{equation}
which guarantees that the width of the singular curve is less than 1, and then choose $\xi_0$ in \eqref{e:center} so that the singular curve lies strictly between the two consecutive integers 0 and 1:
\begin{equation}\label{reg_cond_2}
    \frac{k}{2}\,g(\rho,k)<\xi_0<k-\frac{k}{2}\,g(\rho,k).
\end{equation}
The above conditions guarantee that the KM-like breather \eqref{KM} is regular for all $n\in \mathbb{N}$ and all $t\in \Real$, and an example is plotted in Fig.~3.

\begin{figure}[t!!]
\label{f3}
\centering
\includegraphics[width=9cm,height=6cm]{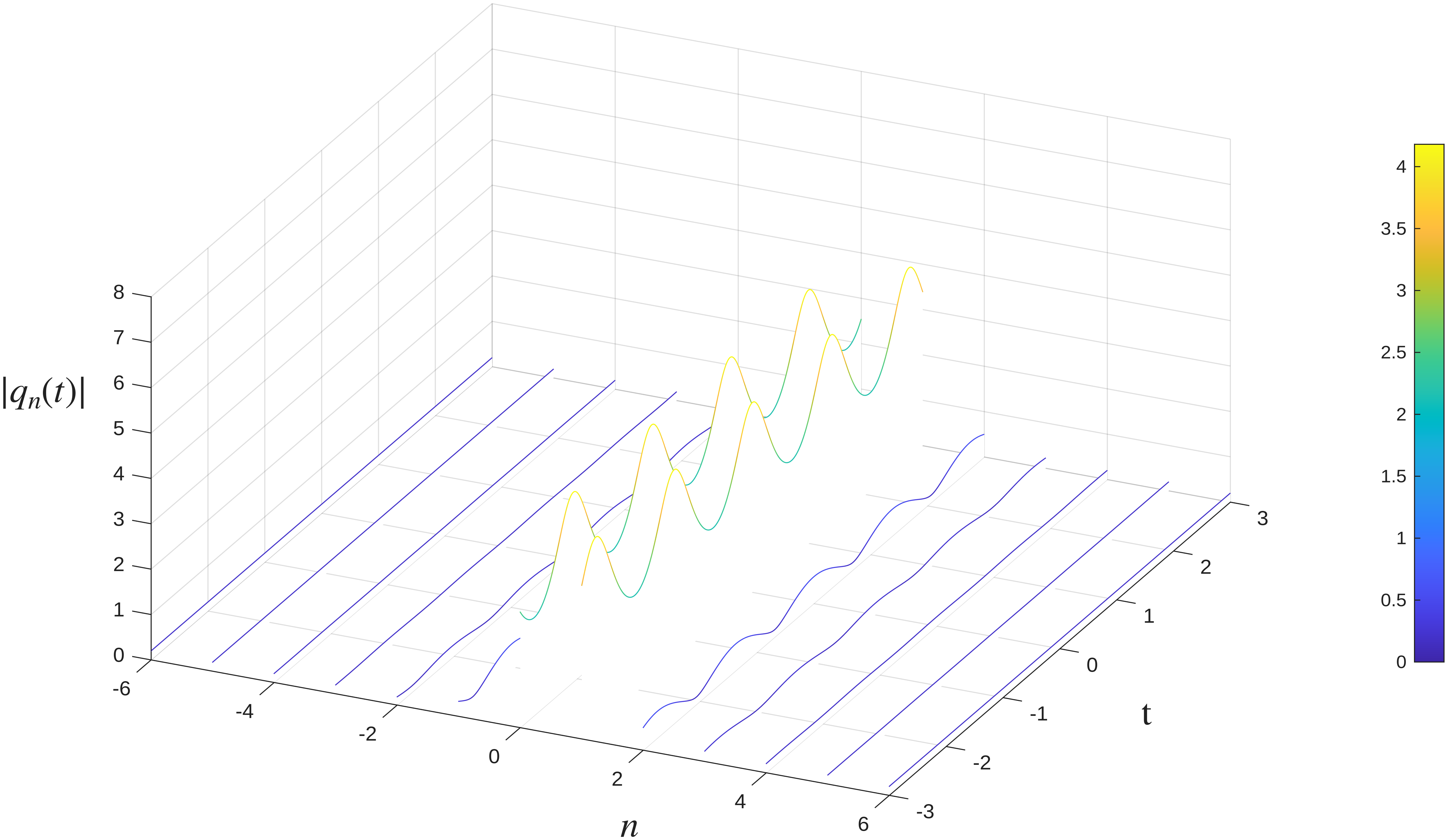}\qquad \qquad
\includegraphics[width=4cm,height=4cm]{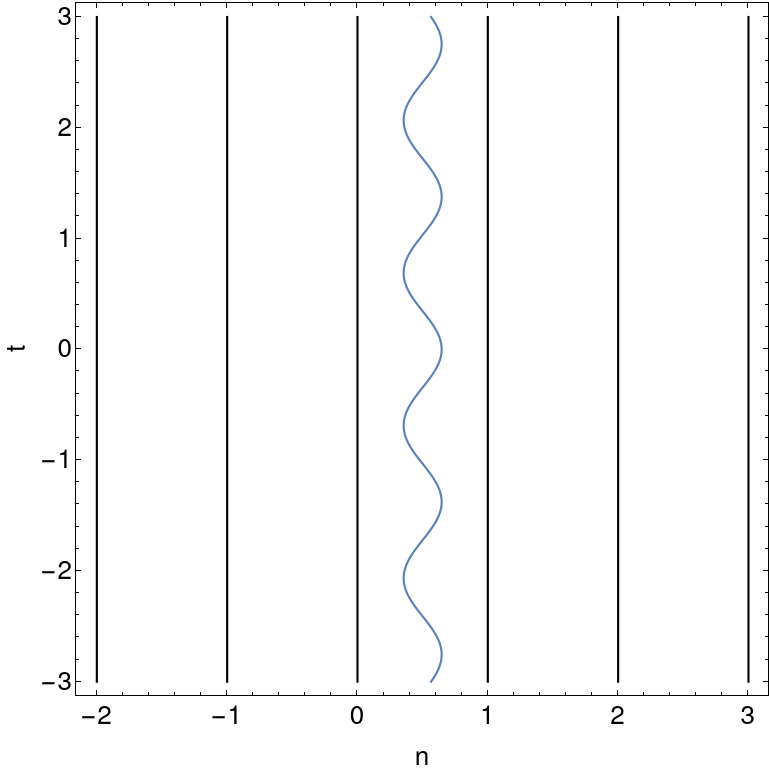}
 \caption{Left: plot of the regular KM breather \eqref{KM} satisfying regularity conditions \eqref{reg_cond_1}, \eqref{reg_cond_2}. Here: $k_0=0$, $\rho=0.2$, $k=1.9$, $\xi_0=\frac{k}{2}$ and $t_0=0$. The existence condition is satisfied since $\delta=5.39>1$. Right: plot below, of the singular curve $n(t) $, defined in \eqref{n_t_singular}, entirely contained within the lattice spacing (black lines).}
\end{figure}

\subsubsection{Solutions with $\delta \to 1$}
For completeness, we discuss below the critical case when $\delta$ approaches one.
When $\delta=1$, both \eqref{theta} and \eqref{phi} give:
$$ \sinh\left(\frac{k}{2}\right)=\frac{\rho}{  \sqrt{1-\rho^2}},
$$ 
and we can introduce: 
$$
\tilde{k}=2\, \textrm{arcsinh}\left(\frac{\rho}{\sqrt{1-\rho^2}}\right),
$$
obtaining two different cases, discussed below.
\begin{itemize}
    \item When the limit is performed from the singular domain we obtain a solution with the following rational-hyperbolic dependence on the variables $n,t$:
      \begin{equation}
        q_n^{(s)}(t)=\rho\, e^{i(k_0 n+\omega_0\,t)}\;\frac{\sqrt{1-\rho^2}\,\sinh(\tilde{k}\,n-4\rho\sin k_0 t)+4\rho^2t\cos k_0-2i}{\sqrt{1-\rho^2}\,\sinh(\tilde{k}\,n-4\rho\sin k_0 t)-4\rho^2\cos k_0t},
    \end{equation}
    where $\omega_0=2\cos k_0(1-\rho^2)$. Setting $k_0=0$ to remove the plane wave yields:
    \begin{equation}
        q_n^{(s)}(t)=\rho\, e^{2i(1-\rho^2)t}\;\frac{\sqrt{1-\rho^2}\,\sinh(\tilde{k}\,n)+4\rho^2t-2i}{\sqrt{1-\rho^2}\,\sinh(\tilde{k}\,n)-4\rho^2t}.
    \end{equation}
    Unlike the periodic case, this solution always develops singularities in finite time.
    \item When the limit is performed from the regular domain we get:
\begin{equation}
    q_n^{(r)}(t) =\rho\, e^{i(k_0 n+\omega_0 t)}\tanh\left(\frac{\tilde{k}\, n +\tilde{\sigma}\, t}{2}\right),
\end{equation}
where $\tilde{\sigma}=2(1-\rho^2)\sinh \tilde{k}\sin k_0$. The latter is a regular dark soliton superimposed to a plane-wave background, and it reduces to a (stationary) black soliton when $k_0=0$ and the plane wave is removed.
\end{itemize}
\subsection{Interactions between a dark soliton and a KM breather}
We now apply the Hirota series procedure to generate a solution describing the interaction between a dark soliton and a breather. As in the previous cases, the zero-th order of the Hirota series is given by the constant solution $    f^{(0)}_n=g^{(0)}_n=1$, and the linear term is the sum of the analog terms of the single breather and dark soliton solutions. We can write the terms in compact form by introducing the phases:
\begin{equation}
\begin{split}
    &z_1(n,t)=k_1\,(n-n_1)-\sigma_r^{(1)}\,t+i\sigma_i^{(1)}\,(t-t_1),\\
    &z_2(n,t)=k_2\,(n-n_2)-\sigma^{(2)}\,t.
    \end{split}
\end{equation}

\begin{equation}
\begin{split}
   & g^{(1)}_n=-e^{z_1(n,t)+2\phi_1}-e^{z_1(n,t)^*-2\phi_1}-e^{z_2(n,t)+2i\theta_2}, \\
   & f^{(1)}_n=e^{z_1(n,t)}+e^{z_1(n,t)^*}+e^{z_2(n,t)},
    \end{split}
\end{equation}
where $z_1(n,t)^*=k_1\,(n-n_1)-\sigma_r^{(1)}\,t-i\sigma_i^{(1)}\,(t-t_1)$ and $(n_1,n_2,t_1)\in\RR$ are free parameters. Since Hirota's equations are linear, the constraints among the parameters are the same as for the individual solutions, namely:
\begin{equation}
    \sigma_r^{(1)}=2(1-\rho^2)\sinh k_1\sin k_0, \qquad \sigma_i^{(1)}=2\rho^2\sinh(2\phi_1)\cos k_0, \qquad  \cosh\phi_1= \frac{  \sqrt{1-\rho^2}}{\rho}\sinh\left(\frac{k_1}{2}\right),
\end{equation}
for the KM breather, obviously with the existence condition for $k_1$: 
$$
\delta_1=\frac{  \sqrt{1-\rho^2}}{\rho}\sinh\left(\frac{k_1}{2}\right)>1.
$$
The analogous constraints for the dark soliton parameters read:
\begin{equation}
     \sigma^{(2)}= 2\rho^2\sin 2\theta_2\cos k_0+2(1-\rho^2)\sinh k_2\sin k_0,\hspace{1cm} \cos\theta_2= \frac{  \sqrt{1-\rho^2}}{\rho}\sinh\left(\frac{k_2}{2}\right),
\end{equation}
along with its existence condition: 
$$
\delta_2=\frac{  \sqrt{1-\rho^2}}{\rho}\sinh\left(\frac{k_2}{2}\right)<1.
$$
The second order of the perturbative series exhibits the interaction among the various modes and can be written as:\\
\begin{equation}
 \begin{split}
   & g^{(2)}_n=a_{11}\,e^{z_1(n,t)+z_1(n,t)^*}+a_{12}\,e^{z_1(n,t)+z_2(n,t)+2\phi_1+2i\theta_2}+a_{12}^*\,e^{z_1(n,t)^*+z_2(n,t)-2\phi_1+2i\theta_2}, \\
   & f^{(2)}_n=e^{z_1(n,t)+z_1(n,t)^*}+e^{z_1(n,t)+z_2(n,t)}+e^{z_1(n,t)^*+z_2(n,t)},
    \end{split}   
\end{equation}
where, as before, 
$$
a_{11}=-\left(\frac{\sinh\phi_1}{\cosh\frac{k_1}{2}}\right)^2,
$$
and the new (complex) interaction coefficient reads:
\begin{equation}\label{a12}
    a_{12}=-\frac{1-\rho^2\cos(2(\theta_2+i\phi_1))-(1-\rho^2)\cosh(k_1-k_2)}{1-\rho^2\cos (2(\theta_2-i\phi_1))-(1-\rho^2)\cosh(k_1+k_2)}.
\end{equation}\\
The third and final term, instead, is given by:\\
\begin{equation}
\begin{split}
&g^{(3)}_n=-f^{(3)}_n=-a_{11}|a_{12}|^2e^{z_1(n,t)+z_1(n,t)^*+z_2(n,t)+2i\theta_2}.
    \end{split}
\end{equation}\\
We can therefore write the final expression as the ratio of the following terms:
\begin{equation}\label{e:KM-Dark}
    q_n(t)=\rho\,e^{i(k_0 n+\omega_0 t)}\;\frac{1+\epsilon\,g_n^{(1)}+\epsilon^2\,g_n^{(2)}+\epsilon^3\,g_n^{(3)}}{1+\epsilon\,f_n^{(1)}+\epsilon^2\,f_n^{(2)}+\epsilon^3\,f_n^{(3)}},
\end{equation}
where again $\omega_0=2(1-\rho^2)\cos  k_0$ and the convergence is therefore guaranteed for every value of  $\epsilon\in\RR$.

\begin{figure}[t!!]
\centering
\includegraphics[height=6cm]{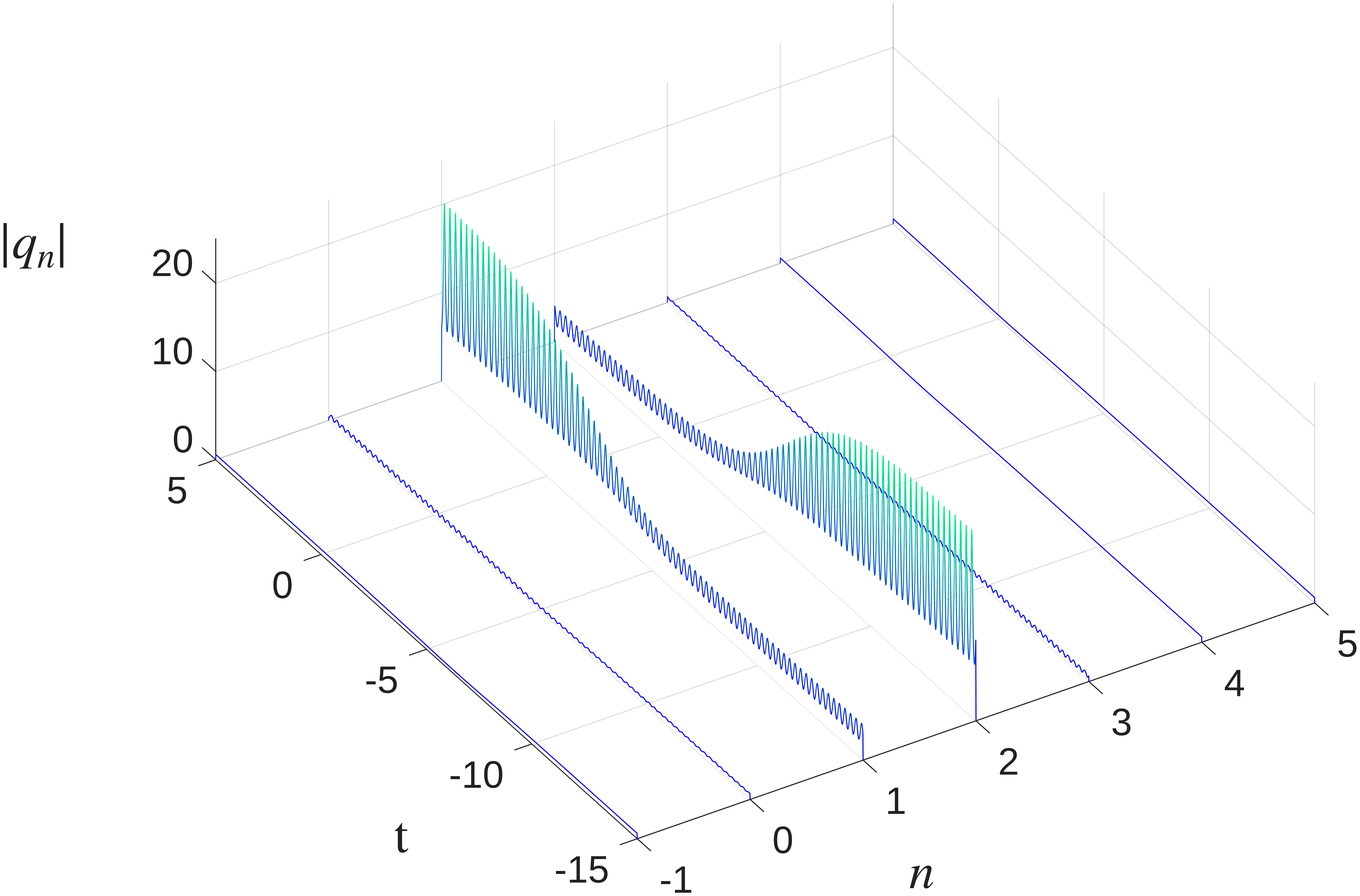} \qquad
\includegraphics[height=5cm]{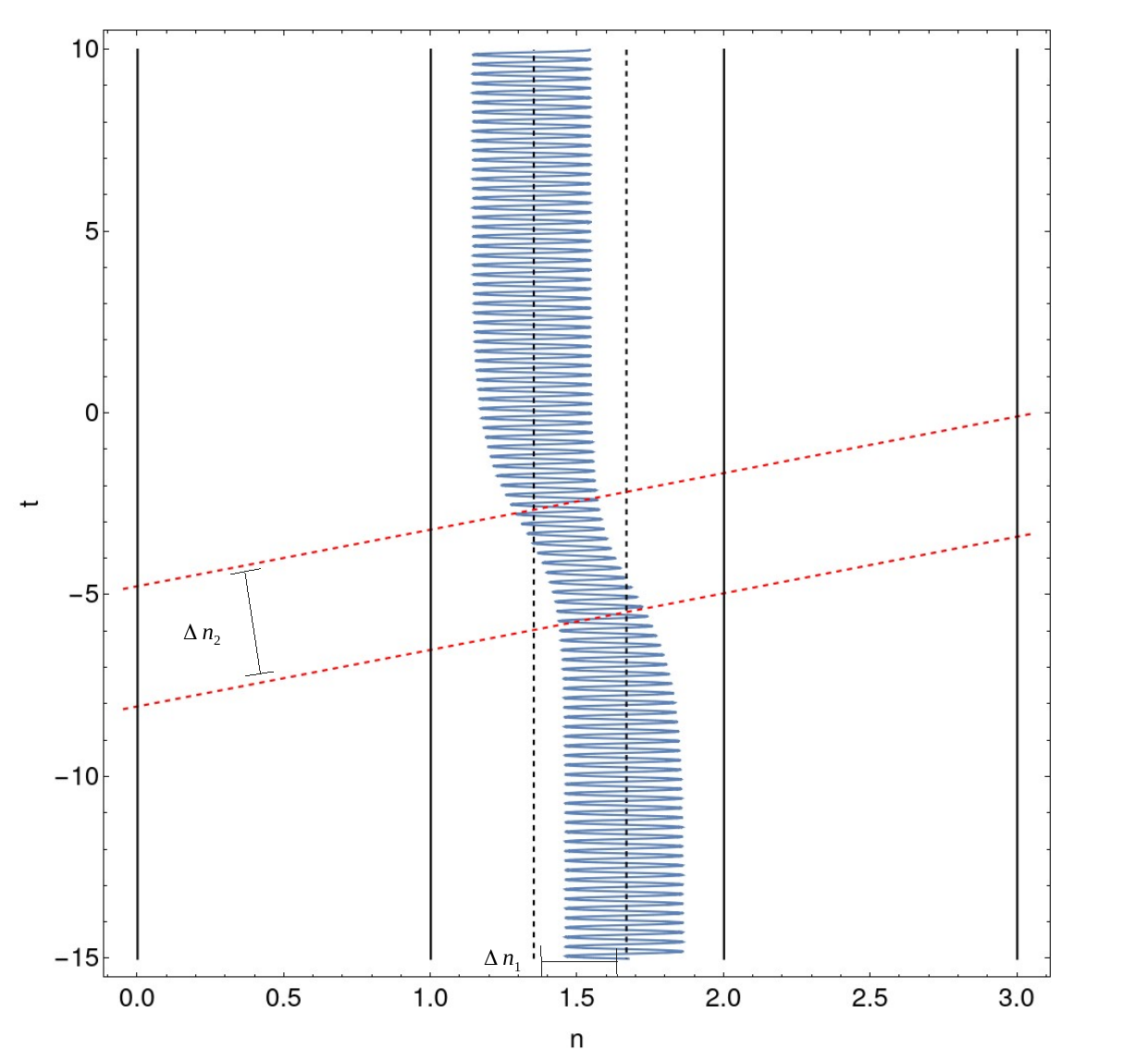}
 \caption{Left: Plot of the modulus of a dark-KM breather solution. Right: Contour plot of the corresponding singular curve, i.e., the curve where the denominator of \eqref{e:KM-Dark} vanishes.  Here: $k_0=0$, $\rho=0.6$, $k_1=3.1, k_2=1.1$, $\epsilon=1$, and position parameters are $n_1=1.4$, $t_1=0$ and $n_2=3$. The existence conditions are fulfilled since $\delta_1=4.14>1$ and $\delta_2=0.77<1$. The black dashed lines indicate the  position shift $\Delta n_1$ of the KM breather due to the interaction with the dark soliton. For completeness, we have also indicated the asymptotic positions of the dark soliton before and after the interaction with dashed red lines. The analytical values are reported in \eqref{km-dark-shift}. }\label{fig:4}
\end{figure}

The solution describes the interaction between a KM breather and a dark soliton; the interaction is elastic, and both components preserve their structure before and after the collision. However, as usually happens, a phase shift occurs in the spatial and temporal positions of the two, which can be derived from a simple asymptotic analysis of the solution. Indeed, one can easily show that in the asymptotic limits $t\rightarrow\pm\infty$ but with fixed $k_1(n-n_1)-\sigma_r^{(1)} t$, i.e., in the reference frame of the KM breather, the phase $z_1$ acquires the complex factor $\log (a_{12})$, defined in \eqref{a12}, for the solution calculated as $t\to +\infty$ with respect to the one as $t\to -\infty$: $z_1\rightarrow z_1+\log(a_{12})$. Similarly, for the dark soliton we obtain a phase difference for the asymptotic regimes $t\to \pm\infty$ given by $z_2\rightarrow z_2+2\log(|a_{12}|)$. The phase shift is described by:
\begin{equation}\label{km-dark-shift}
    \begin{split}
        &n_1\rightarrow n_1+\Delta n_1,  \hspace{1cm}\Delta n_1=\frac{\Re(\log(a_{12}))}{k_1},\\
        &t_1\rightarrow t_1+\Delta t_1,  \hspace{1.3cm}\Delta t_1=\frac{\Im(\log(a_{12}))}{\sigma_i^{(1)}},\\
       & n_2\rightarrow n_2+\Delta n_2,\hspace{1cm}\Delta n_2=2\frac{\log\left(|a_{12}|\right)}{k_2}.
    \end{split}
\end{equation}
The effect of this phase shift is evident in Fig.~\ref{fig:4}, where the deviation of the singular curve from the asymptotic temporal regions due to the interaction with the dark soliton is shown. The regularity condition is analogous to that of the single KM breather case, but in this case the value of the spatial shift must be added to the function $g(\rho,k_1)$ calculated in \eqref{reg_cond_1}, yielding:
 \begin{equation}
     g(\rho,k_1)+|\Re\log(a_{12})|<1.
 \end{equation}

\subsection{Double KM breather}
The construction of the solution describing the interaction between two breathers proceeds in a similar way, but with building blocks of the initial step of the Hirota series given by: 
\begin{equation}
\begin{split}
   & g^{(1)}_n=-e^{z_1(n,t)+2\phi_1}-e^{z_1(n,t)^*-2\phi_1}-e^{z_2(n,t)+2\phi_2}-e^{z_2(n,t)^*-2\phi_2}, \\
   & f^{(1)}_n=e^{z_1(n,t)}+e^{z_1(n,t)^*}+e^{z_2(n,t)}+e^{z_2(n,t)^*},
    \end{split}
\end{equation}
with phases:
\begin{equation}
\begin{split}
    &z_j(n,t)=k_j\,(n-n_j)-\sigma_r^{(j)}\,t+i\sigma_i^{(j)}\,(t-t_j),\\
    &  \sigma_r^{(j)}=2(1-\rho^2)\sinh k_j\sin k_0, \qquad \sigma_i^{(j)}=2\rho^2\sinh(2\phi_j)\cos k_0, \qquad  \cosh\phi_j= \frac{  \sqrt{1-\rho^2}}{\rho}\sinh\left(\frac{k_j}{2}\right),
    \end{split}
\end{equation}
for $j=1,2$ and $(n_1,n_2,t_1,t_2)\in\RR$ are free parameters. The existence conditions are the same for the two modes: 
$$
\delta_j=\frac{  \sqrt{1-\rho^2}}{\rho}\sinh\left(\frac{k_j}{2}\right)> 1, \qquad j=1,2.
$$
The second-order term, namely the first one exhibiting the interactions among the linear modes, is given by:
\begin{equation}
 \begin{split} g^{(2)}_n=a_{11}\,e^{z_1(n,t)+z_1(n,t)^*}+&a_{22}\,e^{z_2(n,t)+z_2(n,t)^*}+a_{12}\,\left(e^{z_1(n,t)+z_2(n,t)+2(\phi_1+\phi_2)}+e^{z_1(n,t)^*+z_2(n,t)^*-2(\phi_1+\phi_2)}\right) \\
   +&a_{21}\,\left(e^{z_1(n,t)+z_2(n,t)^*+2(\phi_1-\phi_2)}+e^{z_1(n,t)^*+z_2(n,t)-2(\phi_1-\phi_2)}\right),\\
f^{(2)}_n=a_{11}\,e^{z_1(n,t)+z_1(n,t)^*}+&a_{22}\,e^{z_2(n,t)+z_2(n,t)^*}+a_{12}\,\left(e^{z_1(n,t)+z_2(n,t)}+e^{z_1(n,t)^*+z_2(n,t)^*}\right) \\
   +&a_{21}\,\left(e^{z_1(n,t)+z_2(n,t)^*}+e^{z_1(n,t)^*+z_2(n,t)}\right),
    \end{split}     
\end{equation}
where:
\begin{equation}
\begin{split}
    & a_{jj}=-\frac{\sinh^2\phi_j}{\cosh^2(k_j/2)} \qquad j=1,2, \\
    &a_{12}=-\frac{1-\rho^2\cosh(2(\phi_1-\phi_2))-(1-\rho^2)\cosh(k_1-k_2)}{1-\rho^2\cosh(2(\phi_1+\phi_2))-(1-\rho^2)\cosh(k_1+k_2)},\\
   &  a_{21}=-\frac{1-\rho^2\cosh(2(\phi_1+\phi_2))-(1-\rho^2)\cosh(k_1-k_2)}{1-\rho^2\cosh(2(\phi_1-\phi_2))-(1-\rho^2)\cosh(k_1+k_2)}.
    \end{split}
\end{equation}
For the third order we obtain:
\begin{equation}
 \begin{split}
    g^{(3)}_n=&-a_{11}a_{12}a_{21}\left(e^{z_1(n,t)+z_1(n,t)^*+z_2(n,t)+2\phi_2}+e^{z_1(n,t)+z_1(n,t)^*+z_2(n,t)^*-2\phi_2}\right)\\
    &-a_{22}a_{12}a_{21}\left(e^{z_1(n,t)+z_2(n,t)+z_2(n,t)^*+2\phi_1}+e^{z_1(n,t)^*+z_2(n,t)+z_2(n,t)^*-2\phi_1}\right),\\
f^{(3)}_n=&a_{11}a_{12}a_{21}\left(e^{z_1(n,t)+z_1(n,t)^*+z_2(n,t)}+e^{z_1(n,t)+z_1(n,t)^*+z_2(n,t)^*}\right)\\
&+a_{22}a_{12}a_{21}\left(e^{z_1(n,t)+z_2(n,t)+z_2(n,t)^*}+e^{z_1(n,t)^*+z_2(n,t)+z_2(n,t)^*}\right)
    \end{split}     
\end{equation}
and the fourth and final term is given by:
\begin{equation}
    g^{(4)}_n=f^{(4)}_n=a_{11}\,a_{22}\,a_{12}^2\,a_{21}^2 \;e^{z_1(n,t)+z_1(n,t)^*+z_2(n,t)+z_2(n,t)^*}.   
\end{equation}
We can recombine the various terms, returning to the original expression of the solution representing the interaction between two KM breathers:
\begin{equation}
\label{e:2KM}
    q_n(t)=\rho\,e^{i(k_0 n+\omega_0 t)}\;\frac{1+\epsilon\,g_n^{(1)}+\epsilon^2\,g_n^{(2)}+\epsilon^3\,g_n^{(3)}+\epsilon^4\,g_n^{(4)}}{1+\epsilon\,f_n^{(1)}+\epsilon^2\,f_n^{(2)}+\epsilon^3\,f_n^{(3)}+\epsilon^4\,f_n^{(4)}}.
\end{equation}
The above solution is singular, in general, but, like for a single KM breather, there exist choices of the breather parameters for which the double KM breather remains regular on the lattice for all times. For instance,  in Fig.~5 a contour plot of the singular curves (i.e., the curves where the denominator of \eqref{e:2KM} vanishes, shows that for specific choices of the parameters the singular curves can be confined within two lattice sites, thus ensuring the regularity of the solution on the lattice. 
%For the parameters in Fig.~5 the 2 breathers are sufficiently far apart that the interaction terms , but it is in principle possible to also obtain regularity conditions for interacting breathers, provided 
\begin{figure}[t!!]
\label{f5}
\centering
\includegraphics[width=10cm,height=6cm]{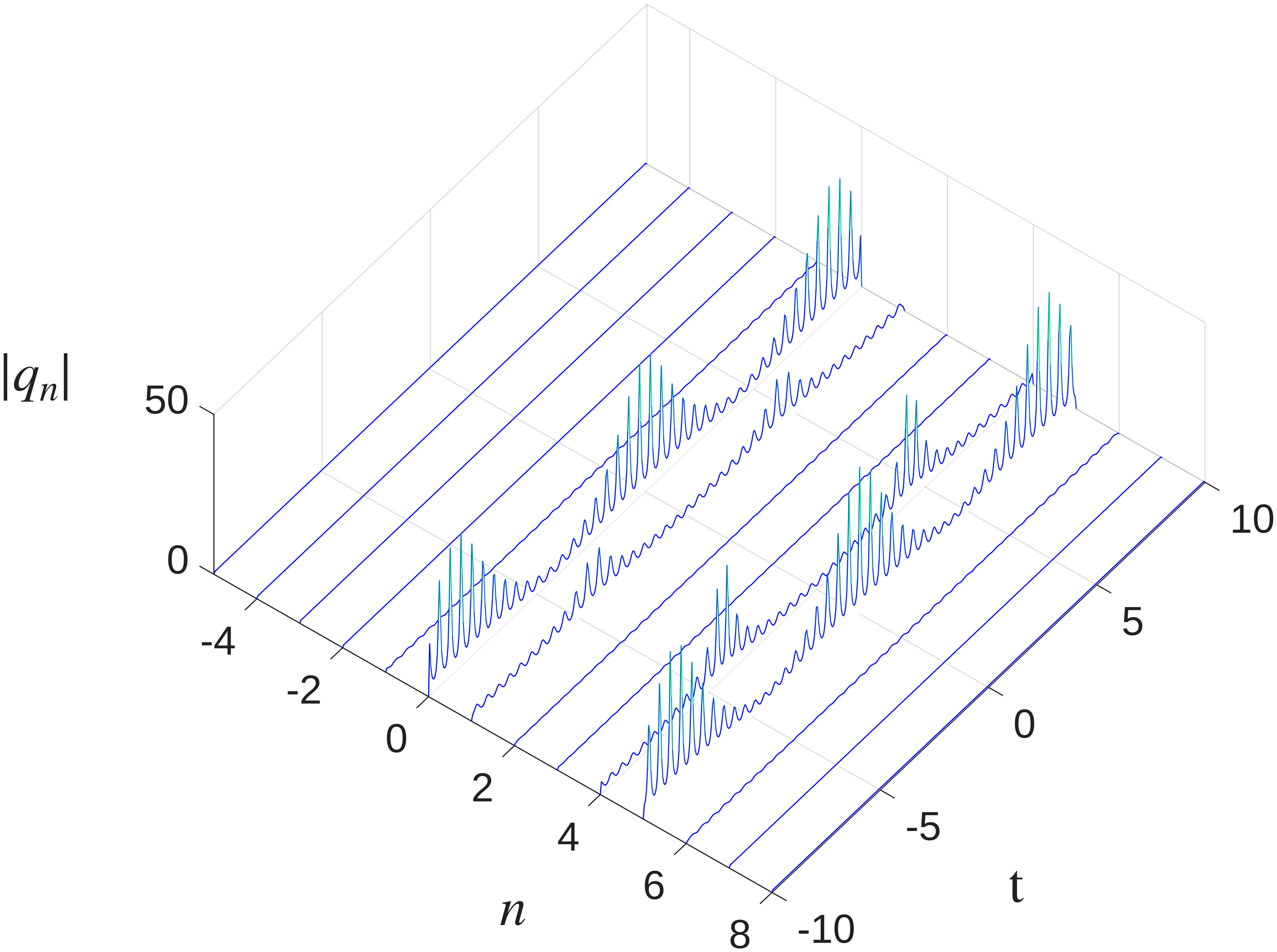}
\includegraphics[width=6cm,height=6cm]{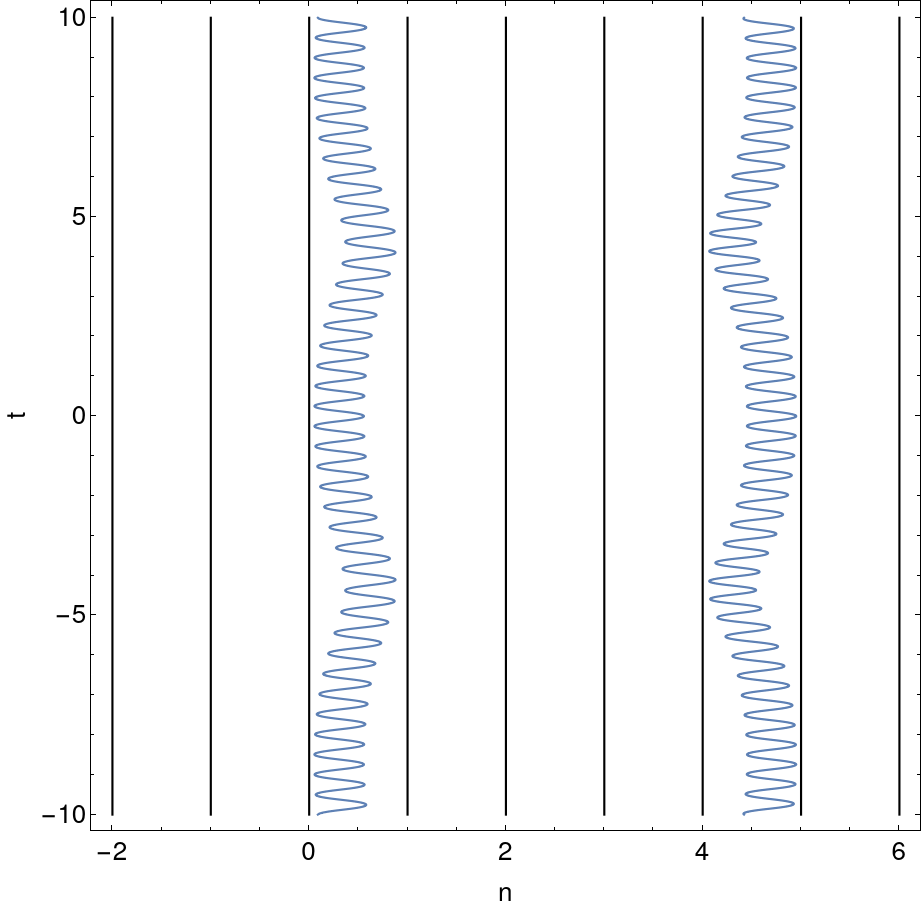}
 \caption{Left: Plot of the modulus of a regular double KM solution. Right: contour plot of the corresponding singular curves, i.e., the curves where the denominator of \eqref{e:2KM} vanishes. Here: $k_0=0$, $\rho=0.6$ and $k_1=3.1$, $k_2=3.15$, $n_1=0.47$, $n_2=0.72$, $\epsilon=1$ and $t_1=t_2=0$.}
\end{figure}

%{\color{red}Aggiungere i plots, contour plots, condizioni di regolarita' (se possibile analitiche, altrimenti dalle figure).}

\section{Integrable discrete Manakov (IDM) system on a small background}

We now consider the defocusing integrable discrete Manakov (IDM) system:
\begin{equation}
\label{e:IDM}
\begin{split}
&  i\partial_tq^{(1)}_n+(q^{(1)}_{n+1}+q^{(1)}_{n-1})\left[1-(|q^{(1)}_n|^2+|q^{(2)}_n|^2)\right]=0,\\[2mm]
&  i\partial_tq^{(2)}_n+(q^{(2)}_{n+1}+q^{(2)}_{n-1})\left[1-(|q^{(1)}_n|^2+|q^{(2)}_n|^2)\right]=0,\\
    \end{split}
\end{equation}
where $\@q_n(t)=(q_n^{(1)}(t),q_n^{(2)}(t))$ is a 2-component complex vector function of $n\in \Z$, the natural generalization to 2 components of the AL system. We first derive the discrete  dark-bright solitons of the IDM system, which reduce to the solutions obtained in \cite{FO17} in the case of a constant background.
We then consider the special case of counter-propagating plane waves, and we obtain two novel classes of
solutions which are periodic in space and homoclinc in time, i.e., discrete Akhmediev-type breathers, corresponding respectively, to the case $\rho_1=\rho_2$ and $\rho_1\ne \rho_2$. Finally, taking the limit as the wavenumber $k_0\to 0$ (i.e., the limit of infinite period), we obtain rational solutions of the defocusing IDM system which are discrete analogs of the ones presented in \cite{Baronio14} for the defocusing Manakov system.

\subsection{Hirota's method for the IDM system}

\subsubsection{Dark-bright solitons}

The generalization of Hirota’s equations to the IDM equation \eqref{e:IDM} starting from the scalar case is fairly straightforward, and it is based on the assumption of a solution written in rational form for both components.
Since the continuous Manakov system also admits dark-bright solitons, we now seek for the analogous discrete solutions by means of the following ansatz for the Hirota equations:
\begin{equation}
\begin{split}
 &   q^{(1)}_n=\rho\,e^{i(k_0 n+\omega_0 t)}\;\frac{g_n^{(1)}}{f_n},\\
 &   q^{(2)}_n=\frac{g_n^{(2)}}{f_n},
    \end{split}
\end{equation}
with dispersion relation given by $\omega_0=2\cos k_0(1-\rho^2)$. By repeating the same approach used in the previous cases, we can identify the Hirota operators:
\begin{subequations}
\label{e:H_IDM}
\begin{align}
H_{1a}[g_n^{(1)},f_n] :=\;&
i\left( (\partial_t g^{(1)}_n)\,f_n - (\partial_t f_n)\,g^{(1)}_n \right) \\
&+ (1-\rho^2)\Big(
e^{i k_0}\, g^{(1)}_{n+1}\,f_{n-1}
+ e^{-i k_0}\, g^{(1)}_{n-1}\,f_{n+1} -2\cos(k_0)\, g^{(1)}_n\,f_n
\Big), \nonumber
\end{align}
\begin{align}
H_{1b}[g^{(2)}_n,f_n] :=\;&
i\left( (\partial_t g^{(2)}_n)\,f_n - (\partial_t f_n)\,g^{(2)}_n \right)+ (1-\rho^2)\left(
g^{(2)}_{n+1}\,f_{n-1} + g^{(2)}_{n-1}\,f_{n+1}
\right),
\end{align}
\begin{align}
H_2[g^{(1)}_n,g^{(2)}_n,f_n] :=\;&
f_n^2
-(1-\rho^2)\,f_{n+1}\,f_{n-1} -\left(
\rho^2\, g^{(1)}_n\,(g^{(1)}_n)^* + g^{(2)}_n\,(g^{(2)}_n)^*
\right).
\end{align}
\end{subequations}
Letting $k_0=0$, we expand the solution in a Taylor series, using the same convention as in the previous case \eqref{Hirota_taylor}. In this case, the anchoring solution will be chosen as:
\begin{equation}
  f^{(0)}_n=  g^{(1,0)}_n=1, \qquad g^{(2,0)}_n=0.
\end{equation}\\
The linear terms will be:
\begin{equation}
\begin{split}
  f^{(1)}_n=  g^{(1,1)}_n=0, \qquad g^{(2,0)}_n=a\, e^{k n-\sigma t+i(l n-\omega t)},
  \end{split}
\end{equation}
with:
\begin{equation}
\sigma=2(1-\rho^2)\sinh k\sin l, \qquad \omega=-2(1-\rho^2)\cosh k\cos l.
\end{equation}
The second order of the Hirota series is given by:
\begin{equation}
    \begin{split}
   &   f^{(2)}_n= c |a|^2  e^{2(k n-\sigma t)}, \\
    &  g^{(1,2)}_n=c |a|^2 e^{2i\theta} e^{2(k n-\sigma t)}, \\
    &g^{(2,2)}_n=0,  
    \end{split}
\end{equation}
and the interaction coefficient $c$ and the phase $\theta$ are given by:
\begin{equation}
\begin{split}
  &  e^{2i\theta}=\left(\frac{\sinh^2k-\sin^2l+2i\sinh k\sin l}{\sinh^2k+\sin^2l}\right)\qquad \Rightarrow \qquad \cos\theta=\frac{\sinh k}{\sqrt{\sinh^2 k+\sin^2 l}},\\[2mm]
   &c=\frac{\sinh^2k+\sin^2l}{2\sinh^2k}\frac{1}{2\rho^2+(1-\rho^2)(\cos 2l-\cosh 2k)}=\frac{1}{4\cos^2\theta}\frac{1}{\left[\rho^2-(1-\rho^2)(\sinh^2k+\sin^2l)\right]}.
    \end{split}
\end{equation}
The resulting expression is therefore given by:
\begin{equation}
\begin{split}
 &   q^{(1)}_n(t)=\rho\,e^{2i(1-\rho^2) t}\;\frac{1-\epsilon^2c |a|^2e^{2i\theta} e^{2(k n-\sigma t)} }{1+\epsilon^2 c |a|^2  e^{2(k n-\sigma t)}},\\
 &   q^{(2)}_n(t)=\frac{\epsilon \,a\, e^{k n-\sigma t+i(l n-\omega t)}}{1+\epsilon^2 c |a|^2  e^{2(k n-\sigma t)} }.
    \end{split}
\end{equation}
In order to simplify the solution, we must distinguish the cases where $c$ is positive or negative, which requires examining the sign of $\left[\rho^2-(1-\rho^2)(\sinh^2k+\sin^2l)\right]\gtreqless 0$:
\begin{itemize}
    \item $c>0$: \textit{regular solutions}.
    \begin{equation}
\begin{split}
 &   q^{(1)}_n(t)=\rho\,e^{2i(1-\rho^2) t}\;\frac{1-\epsilon^2c |a|^2e^{2i\theta} e^{2(k n-\sigma t)} }{1+\epsilon^2 c |a|^2  e^{2(k n-\sigma t)}}
 \equiv \rho\,e^{i\theta}e^{2i(1-\rho^2) t}\;\frac{\frac{e^{-(k n-\sigma t+i\theta)}}{(\epsilon \sqrt{c}|a|)}-(\epsilon \sqrt{c}|a|) e^{k n-\sigma t+i\theta} }{\frac{e^{-(k n-\sigma t)}}{(\epsilon \sqrt{c}|a|)}+(\epsilon \sqrt{c}|a|) e^{(k n-\sigma t)}},\\[3mm]
 &   q^{(2)}_n(t)=\frac{\epsilon \,a\, e^{k n-\sigma t+i(l n-\omega t)}}{1+\epsilon^2 c |a|^2  e^{2(k n-\sigma t)} }=\frac{e^{i\arg a}}{\sqrt{c}}\frac{e^{i(l n-\omega t)}}{\frac{e^{-(k n-\sigma t)}}{(\epsilon \sqrt{c}|a|)}+(\epsilon \sqrt{c}|a|) e^{(k n-\sigma t)} }.
    \end{split}
\end{equation}
After introducing the spatial translation variable $e^{\xi_0}=\epsilon |a|\sqrt{c}$, we can rewrite the solution in compact form:
    \begin{equation}
\begin{split}
    q^{(1)}_n(t)&=
    %=-\rho\,e^{i\theta}e^{2i(1-\rho^2) t}\;\frac{\sinh\left(k n-\sigma t+\zeta+i\theta\right)}{\cosh\left(k n-\sigma t+\zeta\right)}\equiv 
-i\rho\,e^{i\theta}e^{2i(1-\rho^2) t}\;\left[\sin\theta-i\cos\theta\tanh(k n-\sigma t+\xi_0)\right],\\[3mm]
    q^{(2)}_n(t) &=e^{i\arg a}e^{i(l n-\omega t)}\cos\theta \frac{\sqrt{\rho^2-(1-\rho^2)(\sinh^2k+\sin^2l)}}{\cosh\left(k n-\sigma t+\xi_0\right)}.\\
     \end{split}
\end{equation}\\
%\begin{figure}[h!!]
%\label{f21}
%\centering
%\includegraphics[width=12cm,height=6cm]{Manakov_dark-bright.pdf}
% \caption{}
%\end{figure}
\item $c<0$: \textit{singular solution}.\\
By repeating the same steps as in the previous case, while taking into account the sign of c, we can write the solution as:\\
\begin{equation}
\begin{split}\label{dark_bright_singular}
    q^{(1)}_n(t)&=-\rho\,e^{i\theta}e^{2i(1-\rho^2) t}\;\frac{\cosh\left(k n-\sigma t+\xi_0+i\theta\right)}{\sinh\left(k n-\sigma t+\xi_0\right)}=\\[2mm]
 &=-i\rho\,e^{i\theta}e^{2i(1-\rho^2) t}\;\left[\sin\theta-i\frac{\cos\theta}{\tanh(k n-\sigma t+\xi_0)}\right],\\[3mm]
    q^{(2)}_n(t)&=e^{i\arg a}e^{i(l n-\omega t)}\cos\theta \frac{\sqrt{(1-\rho^2)(\sinh^2k+\sin^2l)-\rho^2}}{\sinh\left(k n-\sigma t+\xi_0\right)},
     \end{split}
\end{equation}
where now  $e^{\xi_0}=\epsilon |a|\sqrt{|c|}$. Due to the hyperbolic sine in the denominator, the above solutions in general are singular. However, if $\sigma=0$ (i.e., if the solution is stationary), then choosing the parameters such that $\xi_0/k\in \mathbb{R}\setminus \mathbb{N}$ gives solutions that are regular on the lattice for all times. An example of this solution is given in Fig.~6, as a stationary anti-dark-bright discrete soliton.   
\begin{figure}[t!!]
\label{f6}
\centering
\includegraphics[width=12cm,height=6cm]{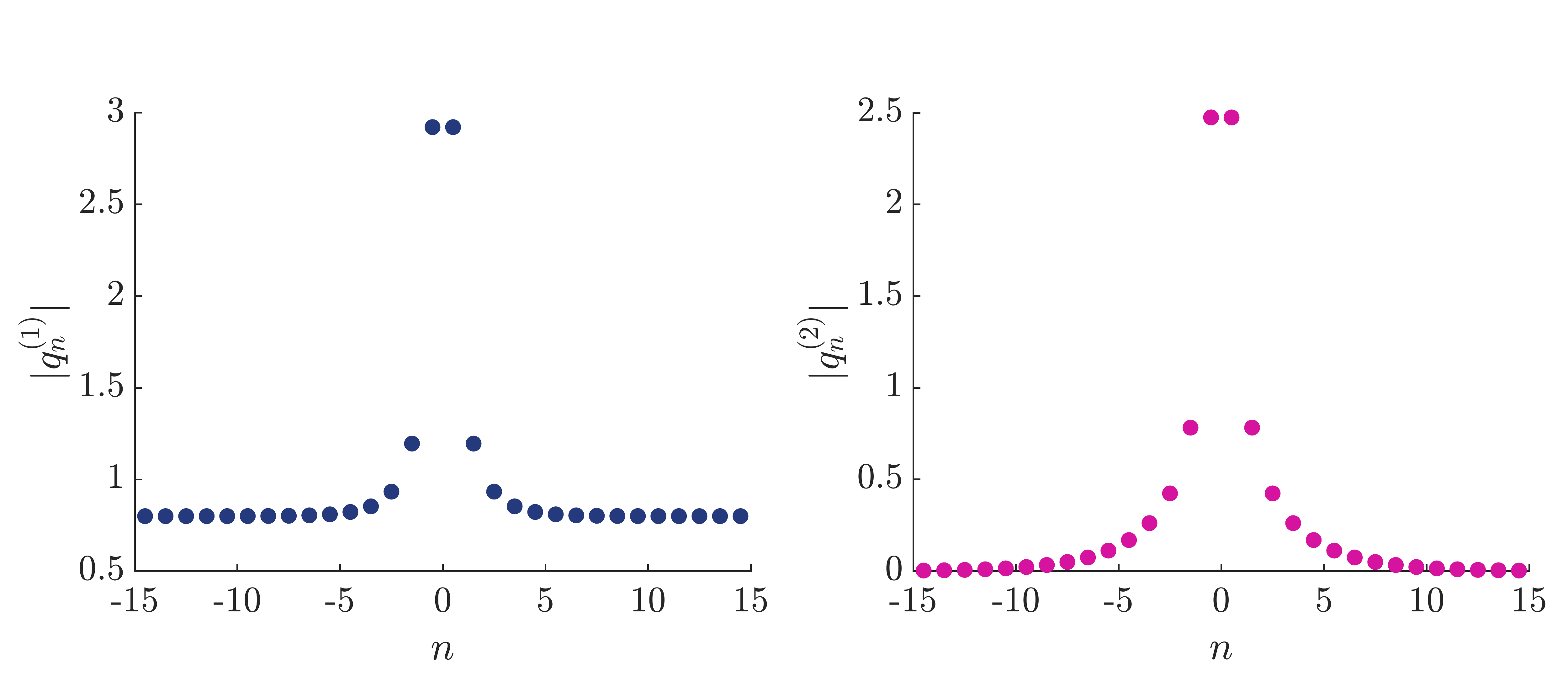}
 \caption{The solution \eqref{dark_bright_singular} in the stationary case $l=0$, for $\rho=0.7$ $k=1.65$ $\epsilon=1$, $a=e^{i\frac{\pi}{4}}$ and $\xi_0=-\frac{k}{2}$.}
\end{figure}
\end{itemize}

\subsubsection{Counter-propagating waves solution}
%$k_0^{(1)}=-k_0^{(2)}$}
The case we want to study in this section is a generalization of the previous one, in which we add the possibility of having a solution that lives on the background in both components. Therefore, one can include the interaction with two independent plane waves, each interacting with one of the field components. We will first derive the Hirota equations for this system in full generality, and subsequently specialize to the specific case of counter-propagating waves, leaving the treatment of the general case to future work
\begin{equation}
\begin{split}
    q^{(1)}_n=\rho_1\,e^{i(k_0^{(1)} n+\omega_0^{(1)} t)}\;\frac{g_n^{(1)}}{f_n},\\
    q^{(2)}_n=\rho_2\,e^{i(k_0^{(2)} n+\omega_0^{(2)} t)}\;\frac{g_n^{(2)}}{f_n},\\
    \end{split}
\end{equation}
where $(g_n^{(1)},g_n^{(2)})\in\mathbb{C}$, $f_n\in\mathbb{R}$ and the dispersion laws are given by 
\begin{equation}
\omega_0^{(j)}\equiv\omega_0\left(k_0^{(j)}\right)=2 \cos k_0^{(j)}\left(1-\tilde{\rho}^2\right). 
\end{equation}
Note that the quantity $\tilde{\rho}^2=\rho_1^2+\rho_2^2$ corresponds to the norm of the vector solution and generalizes the scalar case. The Hirota equations are the following:
{\small\begin{equation}\label{Hirota_counter}
\begin{split}
  &  H_1[g^{(1)}_n,f_n]:=i(\partial_t g^{(1)}_n)f_n-i(\partial_t f_n)g^{(1)}_n +(1-\tilde{\rho}^2)(e^{ik_0^{(1)}} g^{(1)}_{n+1}f_{n-1}+e^{-ik_0^{(1)}} g^{(1)}_{n-1}f_{n+1} -2\cos k_0^{(1)}  \; g^{(1)}_nf_n),\\[2mm]
  &  H_2[g^{(2)}_n,f_n]:=i(\partial_t g^{(2)}_n)f_n-i(\partial_t f_n)g^{(2)}_n +(1-\tilde{\rho}^2)(e^{ik_0^{(2)}} g^{(2)}_{n+1}f_{n-1}+e^{-ik_0^{(2)}} g^{(2)}_{n-1}f_{n+1} -2\cos k_0^{(2)}  \; g^{(2)}_nf_n),\\[2mm]
   & H_3[g_n,f_n]:=f_n^2-\rho_1^2|g^{(1)}_n|^2-\rho_2^2|g^{(2)}_n|^2-(1-\tilde{\rho}^2)f_{n-1}f_{n+1}.
   \end{split}
\end{equation}}
As usual, the solution will be sought in perturbative form by expanding the Hirota equations in powers of the small parameter $\epsilon$:
\begin{equation}\label{Hirota_taylor}
    \begin{split}
       & f_n=f^{(0)}_n+\epsilon f^{(1)}_n+\epsilon^2 f^{(2)}_n+O(\epsilon^3),\\
     &g_n^{(1)}=g^{(1,0)}_n+\epsilon g^{(1,1)}_n+\epsilon^2 g^{(1,2)}_n+O(\epsilon^3),\\
     &g_n^{(2)}=g^{(2,0)}_n+\epsilon g^{(2,1)}_n+\epsilon^2 g^{(2,2)}_n+O(\epsilon^3).
     \end{split}
\end{equation}

We now consider, as previously stated, the special case  of counterpropagating waves, in which the plain-wave vector has opposite wavenumbers in the two components, i.e., we take $k_0^{(1)}=-k_0^{(2)}\equiv k_0$ and $\omega_0^{(2)}=\omega_0^{(1)}\equiv\omega_0$. The initial functions used for the Taylor series expansion of the Hirota solution of \eqref{Hirota_counter} is given by $g^{(1,0)}_n=g^{(2,0)}_n=f^{(0)}_n=1$; then we have for the linear term $(g_n^{(1,1)},g_n^{(2,1)},f^{(1)}_n):$
\begin{equation}
    \begin{split}
       & f^{(1)}_n=e^{\sigma_r\,t+i(2  k n+\sigma_i t)}+e^{\sigma_r\,t-i(2  k n+\sigma_i t)},\\
        & g_n^{(1,1)}=a\,e^{\sigma_r\,t+i(2  k n+\sigma_i t)}+\frac{1}{a^*}e^{\sigma_r\,t-i(2  k n+\sigma_i t)},\\
         & g_n^{(2,1)}=b\,e^{\sigma_r\,t+i(2  k n+\sigma_i t)}+\frac{1}{b^*}e^{\sigma_r\,t-i(2  k n+\sigma_i t)},
    \end{split}
\end{equation}
with the following constraints for the coefficients $(a,b,\sigma_r,\sigma_i)$:
\begin{equation}
\begin{split}
&4(1+b)(1-\rho_1^2-\rho_2^2)\cos k_0\sin^2k+(1-b)(4 \,\Sigma\, \sin k +2(1-\rho_1^2-\rho_2^2)\sin(2 k)\sin k_0)=0,\\[2mm]
&4(1+a)(1-\rho_1^2-\rho_2^2)\cos k_0\sin^2k+(1-a)(4 \,\Sigma\, \sin k-2(1-\rho_1^2-\rho_2^2)\sin(2 k)\sin k_0)=0,\\[2mm]
&-2 ab+(1+a^2)b\rho_1^2+a(1+b^2)\rho_2^2+2ab(1-\rho_1^2-\rho_2^2)\cos(2 k)=0,
\end{split}
\end{equation}
where $i(\sigma_r+i\sigma_i)=4 \,\Sigma\, \sin k $. We can solve the above system with respect to $\Sigma$ obtaining the following relations for the constants a and b:
\begin{equation}
\begin{split}
&a=\frac{\Sigma-(1-\rho_1^2-\rho_2^2)\sin(k_0-k)}{\Sigma-(1-\rho_1^2-\rho_2^2)\sin(k_0+ k)},\\[2mm]
&b=\frac{\Sigma+(1-\rho_1^2-\rho_2^2)\sin(k_0+k)}{\Sigma+(1-\rho_1^2-\rho_2^2)\sin(k_0-k)},
\end{split}
\end{equation}\\
and the parameter $\Sigma$ is a root of $P_4\left(\Sigma\right)=0$, where $P_4$ is the following depressed quartic polynomial:
\begin{equation}\label{quartic}
    P_4(x)=s_0+s_1\, x+s_2\, x^2+x^4,
\end{equation}
where
\begin{equation*}
\begin{split}
&s_0=(1-\rho_1^2-\rho_2^2)^3\left(\cos^2k_0-\cos^2 k\right)\left[\cos^2k_0-(1-\rho_1^2-\rho_2^2)\cos^2k\right],\\[2mm]
&s_1=-2(1-\rho_1^2-\rho_2^2)^2(\rho_1^2-\rho_2^2)\cos^2k_0\sin k_0\cos k,\\[2mm]
&s_2=-(1-\rho_1^2-\rho_2^2)\left[1-(1-\rho_1^2-\rho_2^2)\cos(2 \,k_0)\cos(2\,k)-(\rho_1^2+\rho_2^2)\sin^2k_0\right].
\end{split}    
\end{equation*}
The second (and last) order of the Hirota series is given by:
\begin{equation}
    \begin{split}
       & f^{(2)}_n=\gamma\,e^{2\sigma_r\,t}, \\
        & g_n^{(1,2)}=\frac{a}{a^*}\gamma\,e^{2\sigma_r\,t},\\
         & g_n^{(2,2)}=\frac{b}{b^*}\gamma\,e^{2\sigma_r\,t},
    \end{split}
\end{equation}
where:\\
\begin{equation}
    \gamma=-\frac{\rho_1^2|b|^2\left[1-2\cos(4 k)|a|^2+|a|^4\right]+\rho_2^2|a|^2\left[1-2\cos(4 k)|b|^2+|b|^4\right]-4|ab|^2\sin^2(2 k)}{\rho_1^2(a-a^*)^2|b|^2+\rho_2^2(b-b^*)^2|a|^2}.
\end{equation}
We can therefore write the final expression of the solution as follows:
\begin{equation}\label{counter_sol}
\begin{split}
       & q^{(1)}_n(t)=\rho_1\,e^{i(k_0 n+\omega_0 t)}\;\frac{1+\epsilon\, e^{\sigma_r\,t}\left( a\,e^{i(2  k n+\sigma_i t)}+\frac{1}{a^*}e^{-i(2  k n+\sigma_i t)}\right)+\epsilon^2\frac{a}{a^*}\gamma\,e^{2\sigma_r\,t}}{1+2\epsilon\,e^{\sigma_r\,t}\cos(2  k n+\sigma_i t)+\epsilon^2\gamma\,e^{2\sigma_r\,t}},\\[2mm]
  &  q^{(2)}_n(t)=\rho_2\,e^{i(-k_0 n+\omega_0 t)}\;\frac{1+\epsilon\, e^{\sigma_r\,t}\left( b\,e^{i(2  k n+\sigma_i t)}+\frac{1}{b^*}e^{-i(2  k n+\sigma_i t)}\right)+\epsilon^2\frac{b}{b^*}\gamma\,e^{2\sigma_r\,t}}{1+2\epsilon\,e^{\sigma_r\,t}\cos(2  k n+\sigma_i t)+\epsilon^2\gamma\,e^{2\sigma_r\,t}}.
    \end{split}
\end{equation}
\subsubsection*{Existence conditions}
\begin{figure}[t!!]
\centering
\includegraphics[width=6cm,height=6cm]{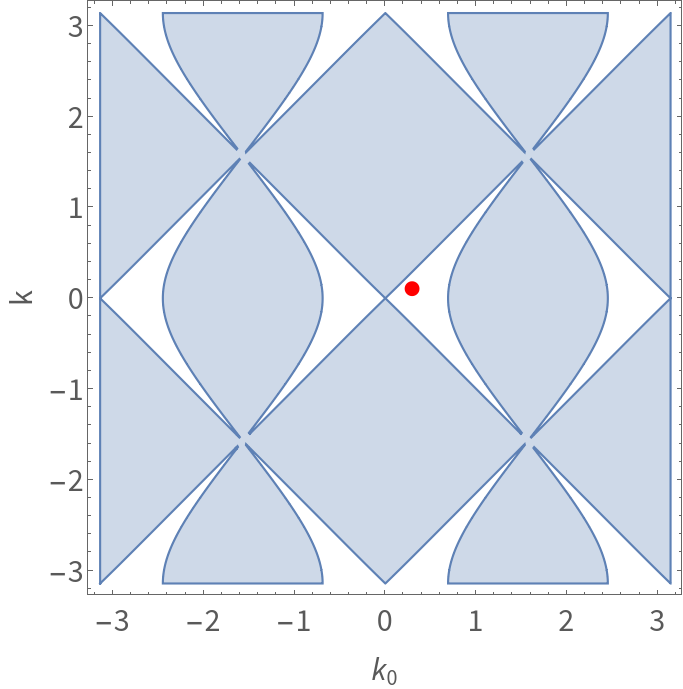}
 \caption{Plot of the existence region (white areas) for the  parameters $\rho_1=0.4$ and $\rho_2=0.5$ in the plain $(k_0,k)\in[-\pi,\pi]$. The red dot corresponds to $(k_0=0.3,k=0.1)$.}
 \label{f7}
\end{figure}

Since we are interested in finding a localized solution on the background, we must require that $\sigma_r$ be different from 0, since the localization properties depend exclusively on this parameter, as can be easily seen from the solution \eqref{counter_sol}. Therefore, one must study the solutions of the quartic polynomial \eqref{quartic}. The quartic polynomial can always be written as the product of two quadratic polynomials, introducing the new constant $\alpha$, as is customary:
\begin{equation*}
    P_4(x)=\left[x^2-\sqrt{2\alpha}\, x+\frac{s_1\sqrt{2\alpha}}{4\alpha}+\frac{s_2}{2}+\alpha\right]\left[x^2+\sqrt{2\alpha}\, x-\frac{s_1\sqrt{2\alpha}}{4\alpha}+\frac{s_2}{2}+\alpha\right]
\end{equation*}
and the constant $\alpha$ is a solution of the cubic equation:
\begin{equation}
    \alpha^3+\alpha^2s_2-\alpha\left(s_0-\frac{s_2^2}{4}\right)-\frac{s_1^2}{8}=0.
\end{equation}
To solve the cubic equation, we shift it by $\alpha=\alpha'-\frac{s_2}{3}$, thus obtaining the depressed cubic:
\begin{equation}
  (  \alpha')^3-\alpha'\left(s_0+\frac{s_2^2}{12}\right)+\frac{1}{216}\left(72s_0 s_2-2s_2^3-27s_1^2\right)=0.
\end{equation}
Then, by introducing the usual quantities $u$ and $v$ we split the equation into the following system:
\begin{equation}\left[
    \begin{split}
       & u^3-v^3=-\frac{1}{216}\left(72s_0 s_2-2s_2^3-27s_1^2\right),\\
       & u\,v=-\frac{1}{3}\left(s_0+\frac{s_2^2}{12}\right),\\
       & u-v=\alpha',
    \end{split}\right.
\end{equation}
whose solution is given by:
\begin{equation}
\begin{split}
    &u^3=-\frac{1}{532}\left(72s_0 s_2-2s_2^3-27s_1^2\right)+\sqrt{\Delta},\\
   &  v^3=\frac{1}{532}\left(72s_0 s_2-2s_2^3-27s_1^2\right)-\sqrt{\Delta},
    \end{split}
\end{equation}
where
\begin{equation}
    \Delta(\rho_1,\rho_2,k,k_0)=\frac{1}{4*216^2}\left(72s_0 s_2-2s_2^3-27s_1^2\right)^2-\frac{1}{27}\left(s_0+\frac{s_2^2}{12}\right)^3.
\end{equation}
In Fig.~\ref{f7} we show the region where, for fixed values of $(\rho_1,\rho_2)$, the function $\Delta(\rho_1,\rho_2,k,k_0)>0$.
\begin{figure}[h!!]
\centering
\includegraphics[width=12cm,height=6cm]{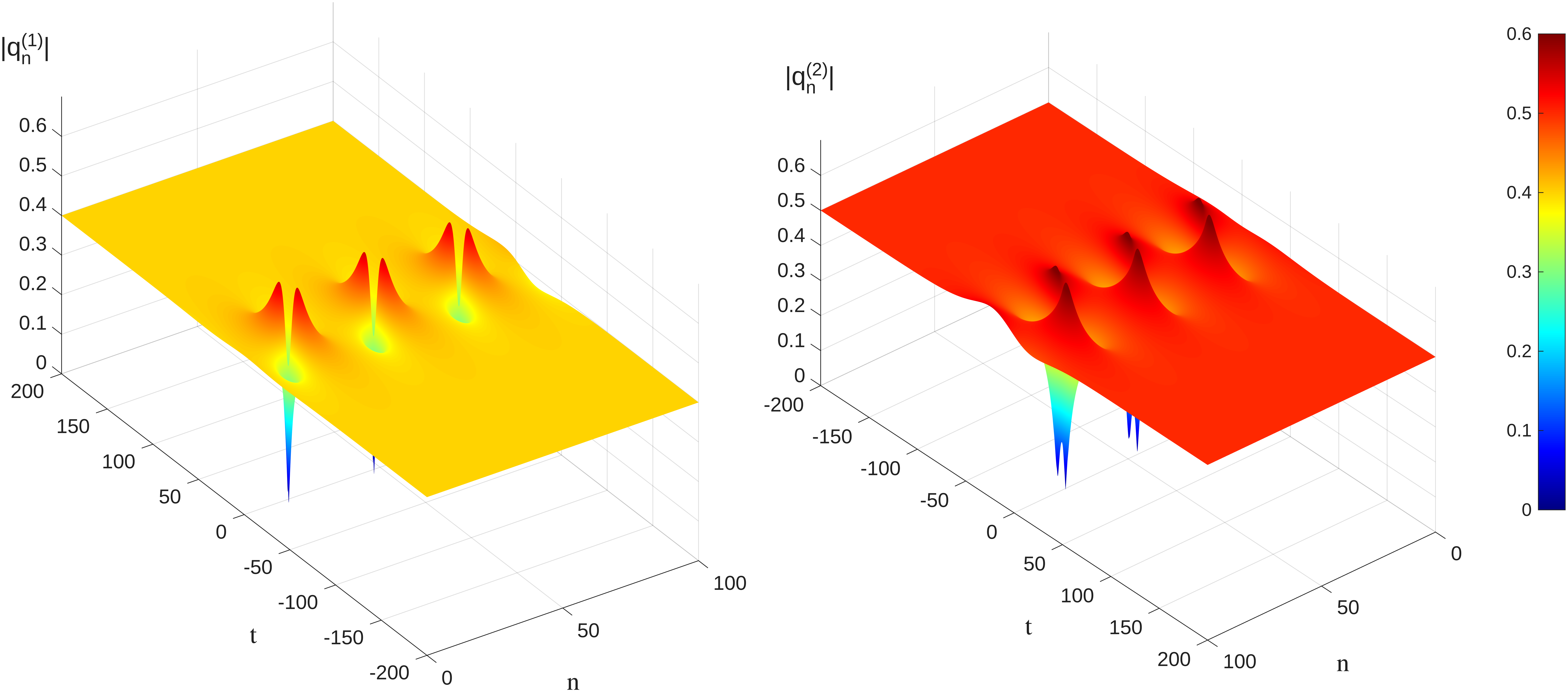}\\[3mm]
\includegraphics[width=8cm,height=4.5cm]{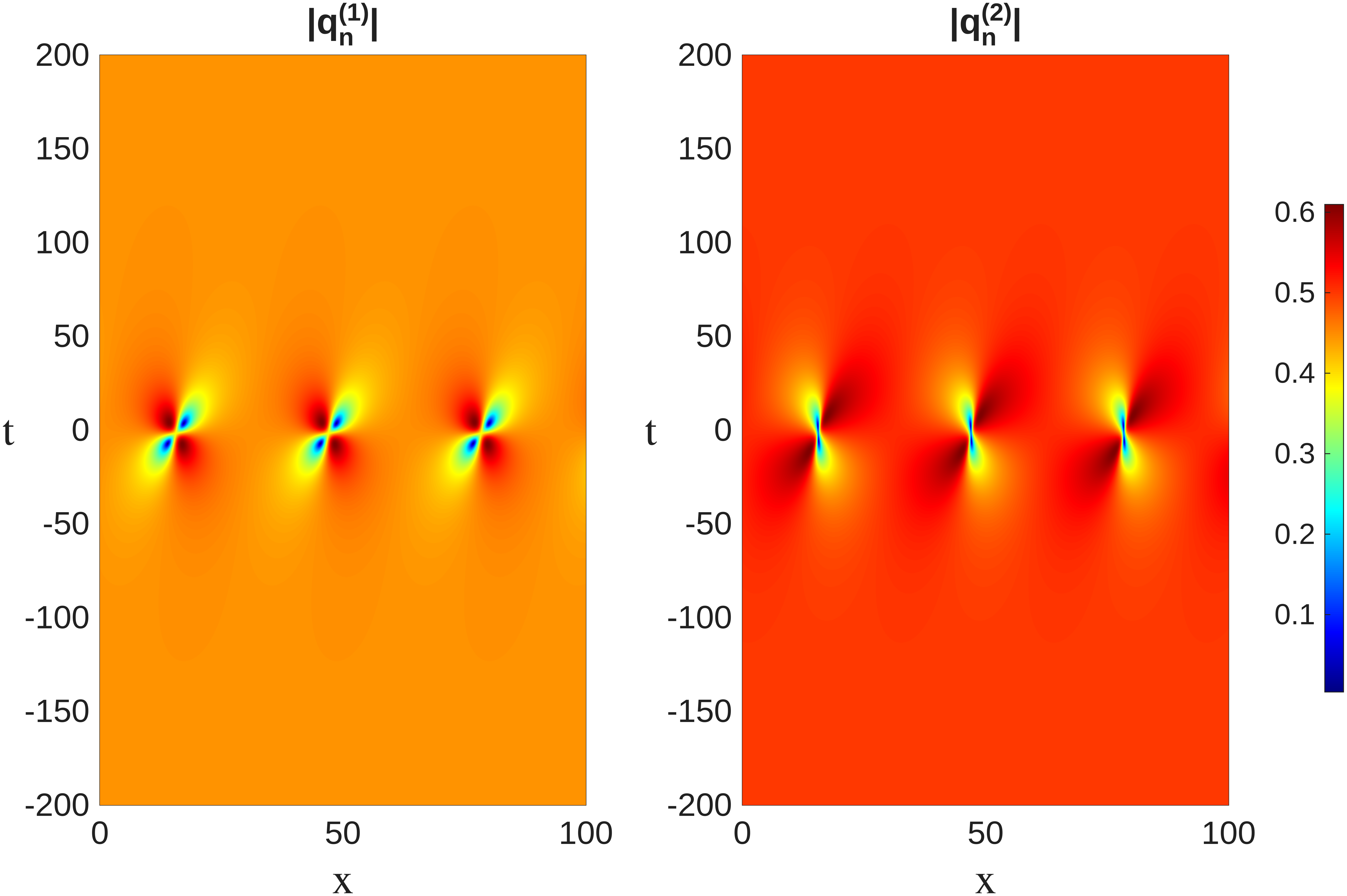}
 \caption{3D plots (top) and density plots (bottom) of the modulii $|q_1^{(1)}|$ and $|q_2^{(1)}|$ of the Akhmediev-type breather solution \eqref{counter_sol} with parameters $k_0=0.3$ $k=0.1$ corresponding to the red dot in Fig.\ref{f7}, and $\rho_1=0.4$, $\rho_2=0.5$.}
\label{f8}
\end{figure}
In Fig.~\ref{f8} we give an example of the Akhmediev-type breather solution \eqref{counter_sol}  corresponding to choices of parameters within the ``existence'' region  in Fig.~\ref{f7}.

\subsection*{Rational solutions}
We can compute the rational limit of the previous formula in the standard way by taking the wave vector $k\rightarrow 0$.  The quantities related to $k$ have the following limits:
\begin{equation}
\begin{split}
 &  \Sigma=  \tilde{\Sigma}+O(k^2), \\
 & a= 1+k\tilde{a}+O(k^2), \\
 & b=  1+k\tilde{b}+O(k^2),\\
&\gamma =1+k^2 \tilde{\gamma}+O(k^4). 
\end{split}
\end{equation}
The solution therefore acquires the following rational form:
\begin{equation}\label{counter_peregrine}
\begin{split}
       & q^{(1)}_n(t)=\rho_1\,e^{i(k_0 n+\omega_0 t)}\;\frac{\tilde{\gamma}-(\tilde{a}^*+2in-t(\tilde{\sigma_r}-i\tilde{\sigma_i}))(\tilde{a}+2in+t(\tilde{\sigma_r}+i\tilde{\sigma_i}))}{\tilde{\gamma}+(2n+t\tilde{\sigma_i})^2+t^2\tilde{\sigma_r}^2},\\[2mm]
  &  q^{(2)}_n(t)=\rho_2\,e^{i(-k_0 n+\omega_0 t)}\;\frac{\tilde{\gamma}-(\tilde{b}^*+2in-t(\tilde{\sigma_r}-i\tilde{\sigma_i}))(\tilde{b}+2in+t(\tilde{\sigma_r}+i\tilde{\sigma_i}))}{\tilde{\gamma}+(2n+t\tilde{\sigma_i})^2+t^2\tilde{\sigma_r}^2}.
    \end{split}
\end{equation}
where $ (\tilde{\sigma_r}+i\tilde{\sigma_i})=-4i  \tilde{\Sigma}$ and $(\tilde{\sigma_r},\tilde{\sigma_i})\in\mathbb{R}^2$.

\begin{figure}[h!!]
\centering
\includegraphics[width=12cm,height=6cm]{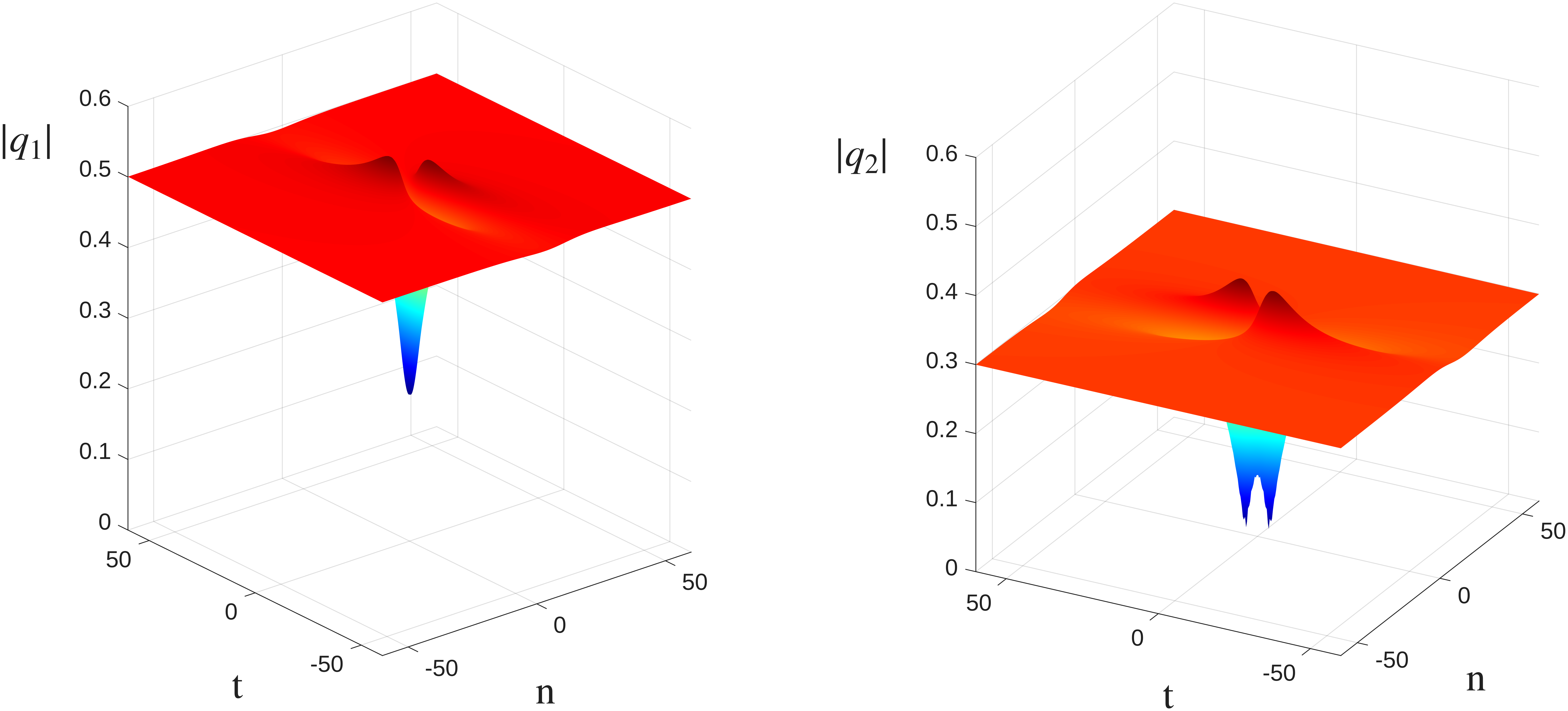}\\[3mm]
\includegraphics[width=11cm,height=5cm]{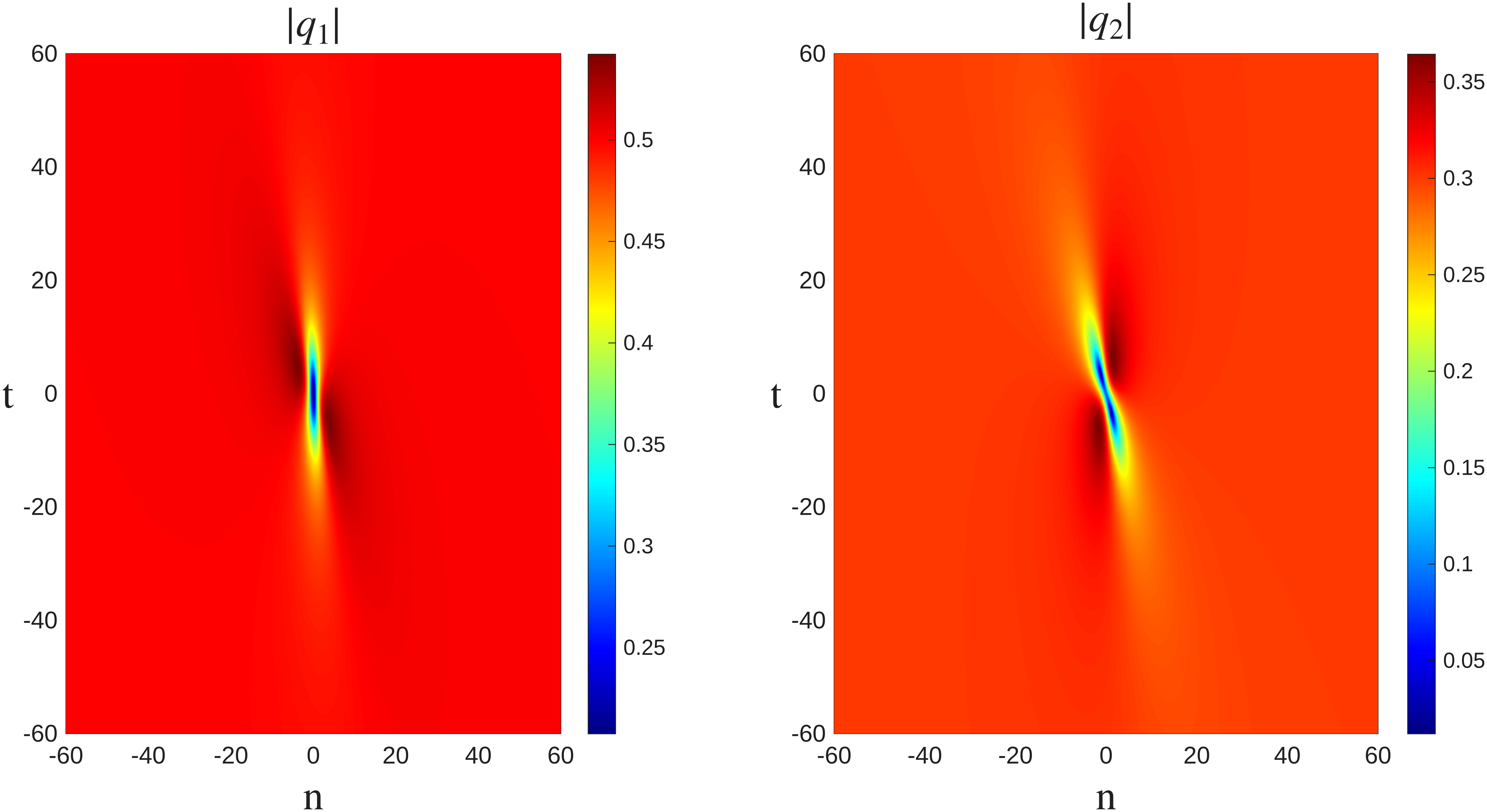}
 \caption{3D plots (top) and density plots (bottom) of the modulii $|q_1^{(1)}|$ and $|q_2^{(1)}|$ of the ratioal solution \eqref{counter_peregrine} with the parameters $\rho_1=0.5$, $\rho_2=0.3$ and $k_0=0.4$. The value of the constants are $\tilde{a}=-3.15-i1.41$, $\tilde{b}=4.06-i3.03$, $\tilde{\sigma}_r=0.57$, $\tilde{\sigma}_i=0.26$ and $\tilde{\gamma}=19.89$.}
\label{f9}
\end{figure}

\section{Conclusions}

In this work we employed Hirota’s bilinear method to construct novel solutions for the scalar and coupled defocusing Ablowitz–Ladik systems in the presence of a small nonzero background. In the scalar case, by establishing a correspondence between the Hirota parameters and the spectral parameters arising in the inverse scattering transform, we clarified the distinction between parameter regimes associated with discrete dark solitons and those generating more exotic (singular, in general) nonlinear structures. In particular, we showed that spectral parameters lying outside the dark-soliton range give rise to novel periodic solutions, including a class of regular breathers which are periodic in time and homoclinic in space, which are the analog of the discrete KM breathes found in the focusing AL equation and in the defocusing AL equation with a large background.  We also discussed the interactions between a dark soliton and a (regular) breather and between two regular breathers. For the coupled AL system, the introduction of counter-propagating plane-wave backgrounds led to new Akhmediev-type discrete breathers which remain regular on the lattice for all times. Finally, in the limit as the wavenumber approaches zero (i.e., as the period approaches infinity) novel rational solutions of the coupled system are obtained. These results further illustrate the rich structure of coherent states supported by integrable discrete systems with nonzero boundary conditions and demonstrate the effectiveness of combining bilinear techniques with spectral considerations.

Several interesting open problems naturally emerge from the present study. A natural follow-up problem is understanding the spectral properties of the novel breathers and rational solutions of the coupled AL lattice, with an analytic characterization of their existence region from a spectral point of view, and a comparison to the recent results for the continuous Manakov equations \cite{Akh24}.

Another important direction concerns the construction of vector rogue-waves arising on standing periodic waves, generalizing the results in \cite{Pel24,Bar26}, where the interaction between localization and periodic modulation revealed a rich class of transient extreme-wave phenomena in the scalar AL equation. This would provide a discrete vector analogue of rogue-wave dynamics on nontrivial backgrounds and could uncover new mechanisms for spatiotemporal localization in integrable lattices. Another challenging problem is the derivation of higher-order rogue waves and their interaction patterns in the presence of counter-propagating discrete plane waves.

A third major open problem concerns the stability of the breathers and rogue waves derived here. While the orbital stability theory for breather solutions of coupled nonlinear Schr\"odinger equations has recently been developed in \cite{Pel26}, analogous results for discrete integrable systems remain largely unexplored. In particular, it would be important to determine whether the regular Akhmediev-type discrete breathers constructed in this paper possess orbital or spectral stability properties under small perturbations. Such an analysis is expected to be considerably more delicate because of the interplay between discreteness, nonzero background states, and the oscillatory structure of the breathers. More broadly, understanding the long-time dynamics and nonlinear stability of coherent structures in coupled discrete systems with finite-density backgrounds remains an open and largely unexplored area of research.

\section*{Acknowledgements}
The authors gratefully acknowledge partial support for this work from the NSF, under grant DMS-2406626. BP acknolwedges the Fulbright Foundation in Greece and the Fulbright program, and the Mathematics Department of the University of Ioannina, Greece, for the kind hospitality during the completion of this work. 

\section*{Conflict of interest}
The authors declare that they have no conflict of interest.

\end{document}